\documentclass[%
 reprint,
 superscriptaddress,
 amsmath,amssymb,
 aps,
]{revtex4-2}
\usepackage{xcolor}
\usepackage{graphicx}
\usepackage{dcolumn}
\usepackage{bm}
\usepackage{soul}

\begin{document}
\preprint{APS/123-QED}

\title{Thermally-controlled magnetooptical metasurface for tunable Faraday rotation}

\author{P. V. Golovko}
\email[]{p.golovko@rqc.ru}
\affiliation{Russian Quantum Center, 121205 Moscow, Russia}
\affiliation{Moscow Institute of Physics and Technology (MIPT), 141700 Dolgoprudny, Russia}

\author{D. O. Ignatyeva}
\affiliation{Russian Quantum Center, 121205 Moscow, Russia}
\affiliation{Faculty of Physics, M.V. Lomonosov Moscow State University, Moscow, Russia}

\author{S. Xia}
\affiliation{National Engineering Research Center of Electromagnetic Radiation Control Materials, University of
Electronic Science and Technology of China, Chengdu 610054, China}
\affiliation{State Key Laboratory of Electronic Thin-Films and Integrated Devices, University of Electronic Science and
Technology of China, Chengdu 610054, China}

\author{P. E. Zimnyakova}
\affiliation{Russian Quantum Center, 121205 Moscow, Russia}

\author{L. Bi}
\affiliation{National Engineering Research Center of Electromagnetic Radiation Control Materials, University of
Electronic Science and Technology of China, Chengdu 610054, China}
\affiliation{State Key Laboratory of Electronic Thin-Films and Integrated Devices, University of Electronic Science and
Technology of China, Chengdu 610054, China}

\author{V.I. Belotelov}
\affiliation{Russian Quantum Center, 121205 Moscow, Russia}
\affiliation{Faculty of Physics, M.V. Lomonosov Moscow State University, Moscow, Russia}

\date{\today}

\begin{abstract}
We present novel type of tunable magneto-optical metasurfaces performing Faraday rotation, the sign and value of which are not fixed after the structure fabrication but can be tuned in a wide range via heating of the metasurface. We demonstrate both experimentally and theoretically that the Faraday rotation angle is enhanced in the vicinity of the magnetodipole and electrodipole Mie resonances and can be changed in a wide range from -0.3 degrees to +0.1 degrees for the same metasurface at fixed wavelength of incident light under the temperature changes from the 294~K to 488~K.  Such thermal heating can be performed by an external control laser. As laser radiation can be focused at the spots of $~\sim 1\mu$m diameter, the magneto-optical response can also be tuned locally. Thus one may obtain the inhomogeneous magnetooptically-induced polarization rotation distributions across the metasurface by the creation of the laser beam patterns with the desired intensity profiles. Another possibility opened by the proposed merasurface is self-modulation of polarization of laser light performed depending on its intensity.
\end{abstract}
\maketitle

\section{Introduction}

Magneto-optical effects and nanostructured materials are crucial in modern technical applications due to their unique properties and high functionality~\cite{kimel20222022}. Advancements in nanophotonics open up opportunities for advanced technical systems and efficient information processing~\cite{smigaj2010magneto,inoue2013magnetophotonics,karki2017thin, ho2018switchable}. Various types of magnetic structures capable of enhancing magneto-optical effects have been proposed recent years~\cite{maccaferri2016anisotropic,fan2019magneto,guo2018significant,xia2022enhancement,barsukova2019enhanced,christofi2018giant,sadatgol2016enhanced,royer2020enhancement,bsawmaii2022magnetic,zograf2021all, testa2021caging}. Different structures can enhance the magneto-optical Faraday effect or the Kerr effects due to the excitation of waveguide modes~\cite{kolmychek2015magneto,ignatyeva2020all,voronov2020magneto,bsawmaii2020longitudinal,pomozov2022resonant}, Mie-type modes~\cite{barsukova2017magneto,chernyak2024faraday}, eigenmodes of parts of the metasurfaces~\cite{zimnyakova2021two}, etc. For all of such structures, the possibility to change the working wavelength is achieved only by changing the parameters of the sample during the production process. Changing the geometric parameters of the structure (period, thickness of various layers, etc.) change the wavelength at which the mode is excited, thereby changing the wavelength at which a significant enhancement of the magneto-optical effect is observed~\cite{krichevsky2021silicon}. 

The magneto-optical response of the magnetic metasurface itself could be tuned via external stimulus, for example~\cite{chang2012temperature, zhang2019electrically, li2020tunable}. The most simple for realization way of such a control is thermal heating. For example, in Ref.~\cite{henriksson2024probing}, it was shown that in hyperbolic meta-antennas, heating enhances electron-phonon scattering, leading to increased damping, a red shift of the nonradiative resonance, and a reduction in the absorption cross-section, which can be leveraged for dynamic tuning of optical and magneto-optical properties.

In this paper, the concept of metasurface on-the-fly tunability is elaborated based on the sensitivity of its optical and, consequently, magneto-optical properties on the parameters of the materials forming it. We experimentally and theoretically investigate a metasurface formed by Si nanocylinders supporting Mie resonances on the top of the Dy:CeYIG magneto-optical film. We show that thermal heating results in the shift of the resonant positions and consequent variations of the magneto-optical response. Sharp Fano shape of the magneto-optical resonance allows for the significant thermal changes of both sign and value of the Faraday effect. We discuss possible applications of the considered metasurface for the tunable light magneto-optical polarization rotation via an external control laser, self-modulation of laser polarization, and sensing.

\begin{figure*}[htb]
\centering
(a)~~~~~~~~~~~~~~~~~~~~~~~~~~~~~~~~~~~~~~~~~~~~~~~~~~~~~~~~~~~~~~~~~~~~~~~~~~~~~~~~~~~~~~~~~(b)\\
\includegraphics[width=0.45\linewidth]{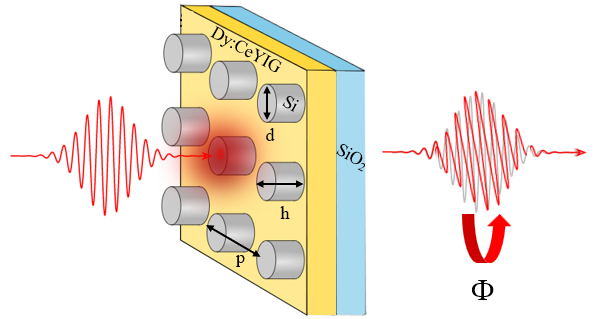}
\includegraphics[width=0.3\linewidth]{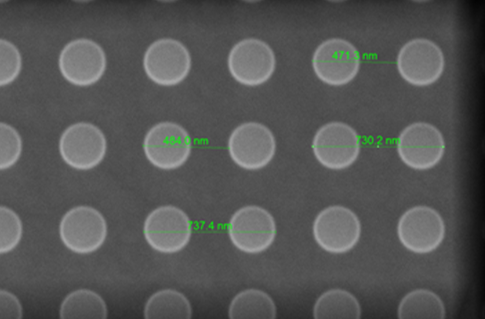}\\
(c)~~~~~~~~~~~~~~~~~~~~~~~~~~~~~~~~~~~~~~~~~~~~~~~~~~~~~~~~~~~~~~~~~~~~~~~~~~~~~~~~~~~~~~~~~(d)\\
~~~~~~~~~experiment~~~~~~~~~~~~~~~~~~~~~~~~~~~~~~~~~~~~~simulations~~~~~~~~~\\
\includegraphics[width=0.32\linewidth]{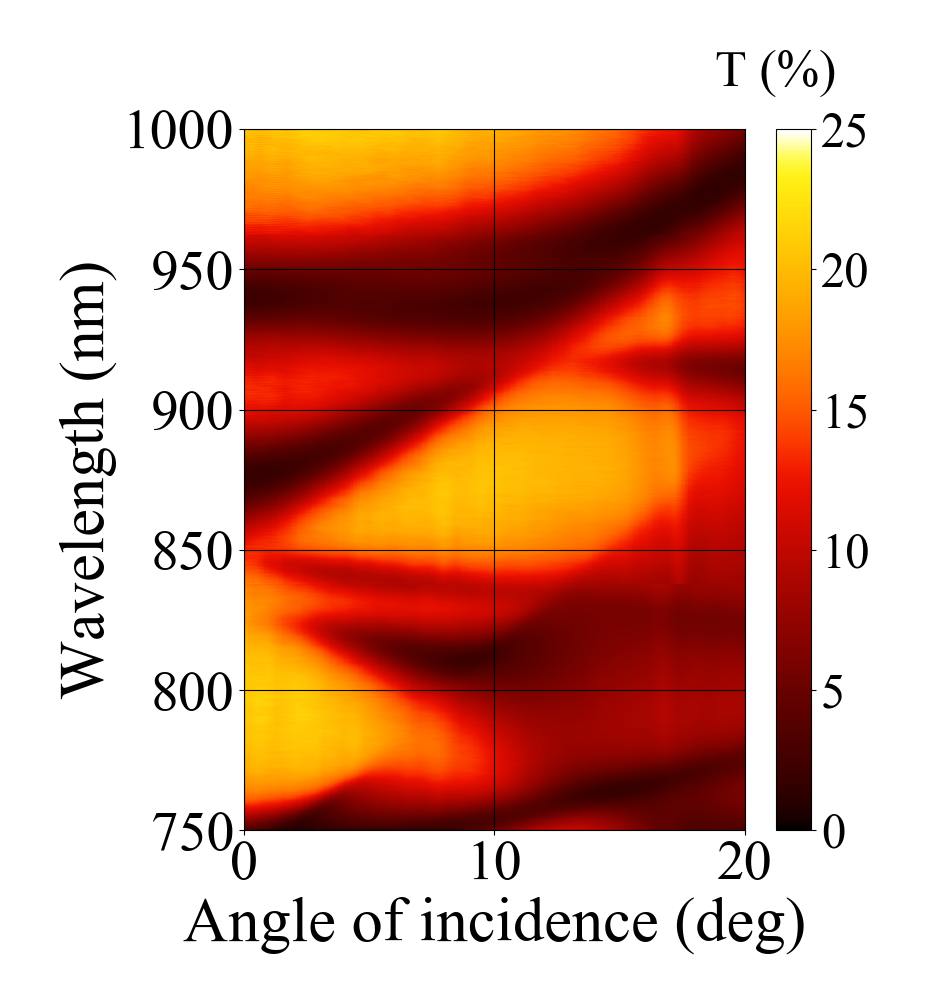}
\includegraphics[width=0.32\linewidth]{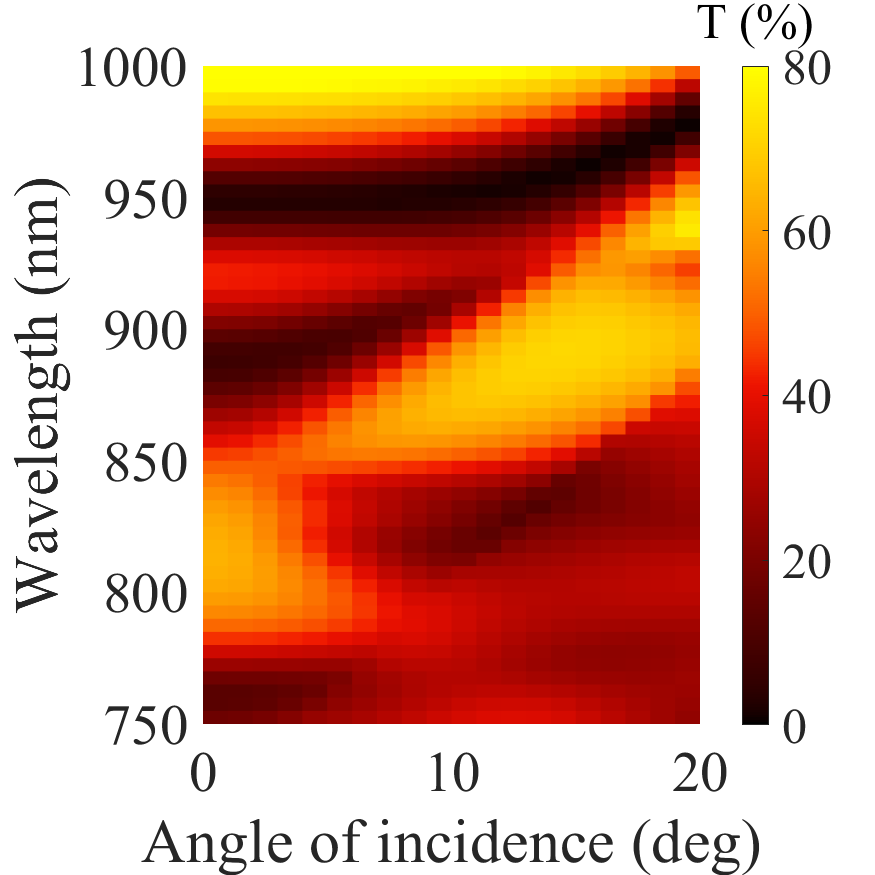}
(e)~~~~~~~~~MD-mode~~~~~~~~~~~~~~~~~~~~~~~~~~~~~~~~~~~~~~~~ED-mode~~~~~~~~~(f)\\
\includegraphics[width=0.35\linewidth]{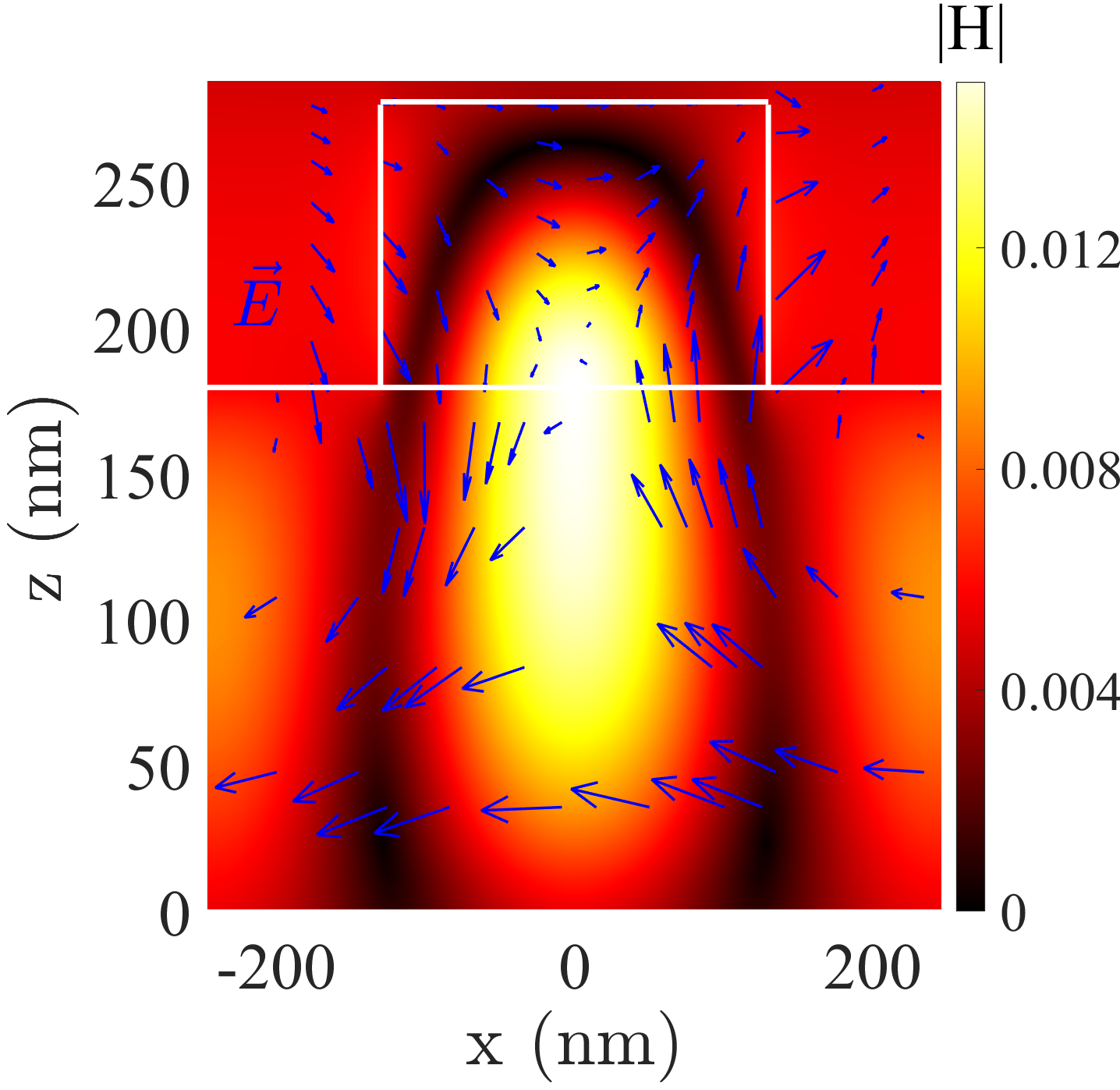}
\includegraphics[width=0.35\linewidth]{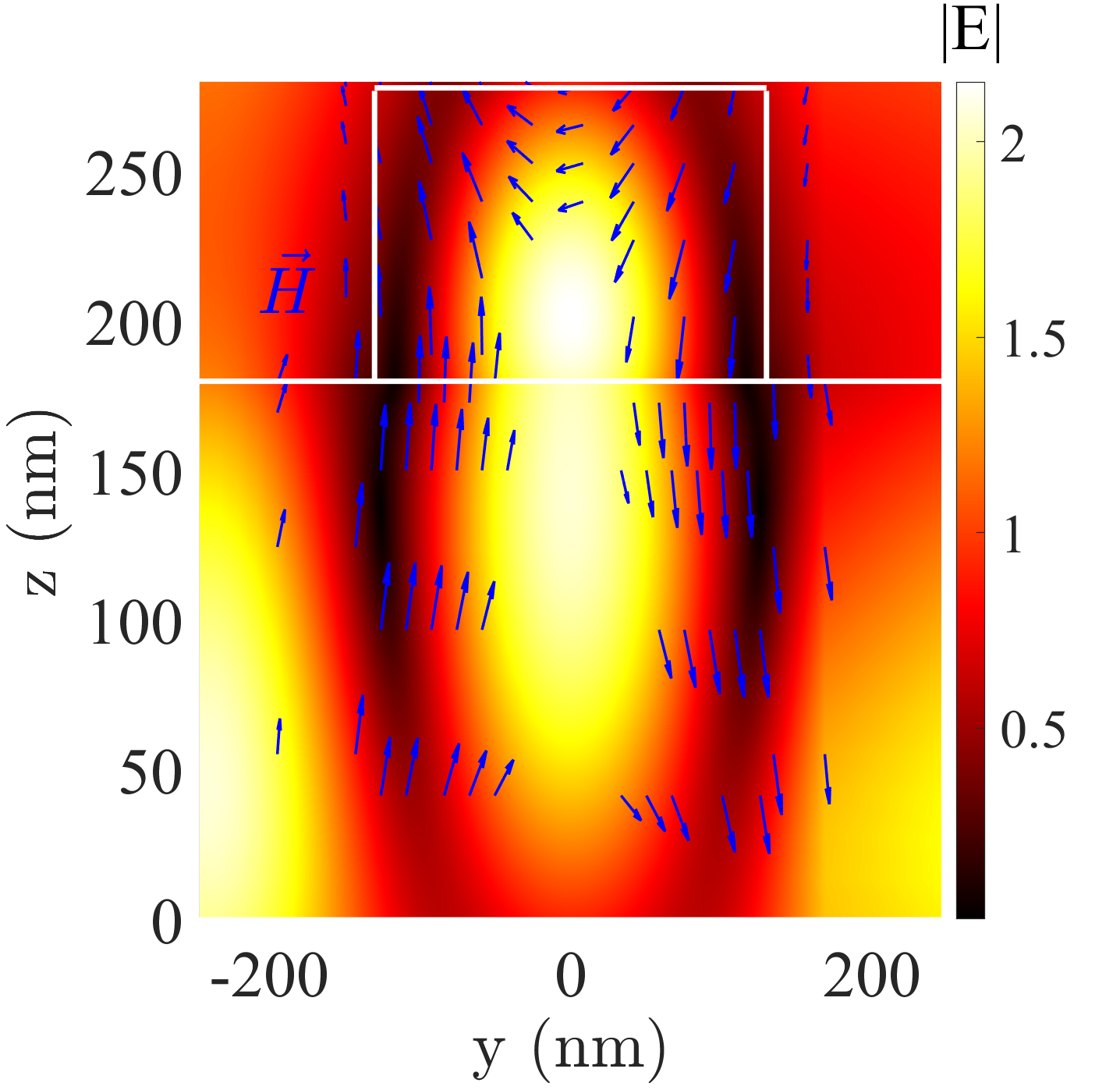}
 \caption{Principal scheme (a) and SEM image (b) of a hybrid all-dielectric magnetic metasurface, the experimental (c) and numerical (d) structure transmittance spectra. (e,f) Electromagnetic near-field distributions at (e) $\lambda=875$~nm and (f) $\lambda=952$~nm wavelengths in $xz$ and $yz$ planes, correspondingly. $|H|$ and $|E|$ are shown by color, and $\mathbf{E}$ and $\mathbf{H}$ vectors are shown by blue arrows in (e) and (f), correspondingly. The incident light is linearly polarized and has $E_x$ and $H_y$ components. Si cylinder and magnetic film layer are shown by the white line.}
    \label{Fig: Modes in metasurface}
\end{figure*}

\section{Results}

We consider a hybrid all-dielectric magnetic metasurface containing a smooth Dy:CeYIG layer of 200 nm thickness and Si nanodisk pattern with $P=492$ nm period on its top. The height of the Si disk is 117 nm, and the radius of the disk is 140 nm. The metasurface is schematically shown in Fig.\ref{Fig: Modes in metasurface}a. Due to the high refractive index of nanodisks, such a metasurface supports excitation of Mie resonances.

\begin{figure*}[htb]
 (a)~~~~~~~~~~~~~~~~~~~~~~~~~~~~~~~~~~~~~~~~~~~~~~~~~~~~~~~~~~~~~~~~~~~~~~~~~(b)\\
\includegraphics[width=0.45\linewidth]{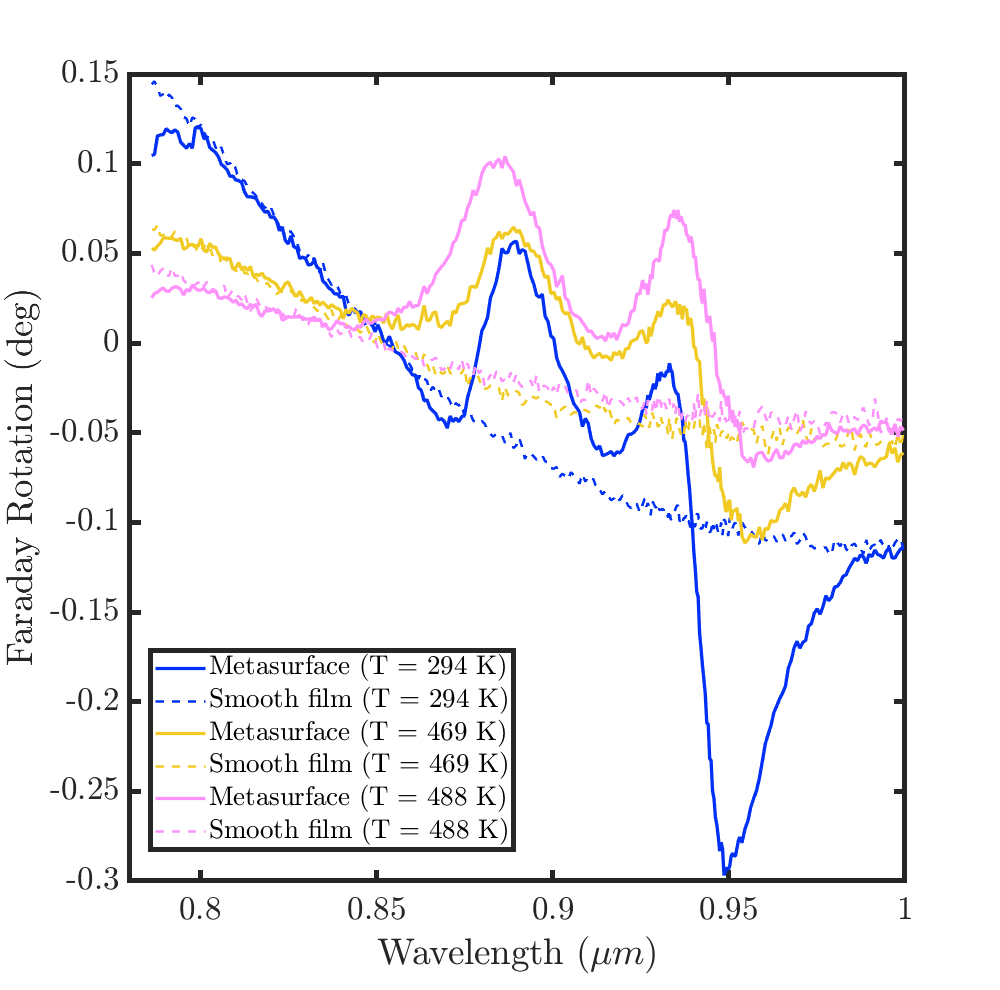}
\includegraphics[width=0.45\linewidth]{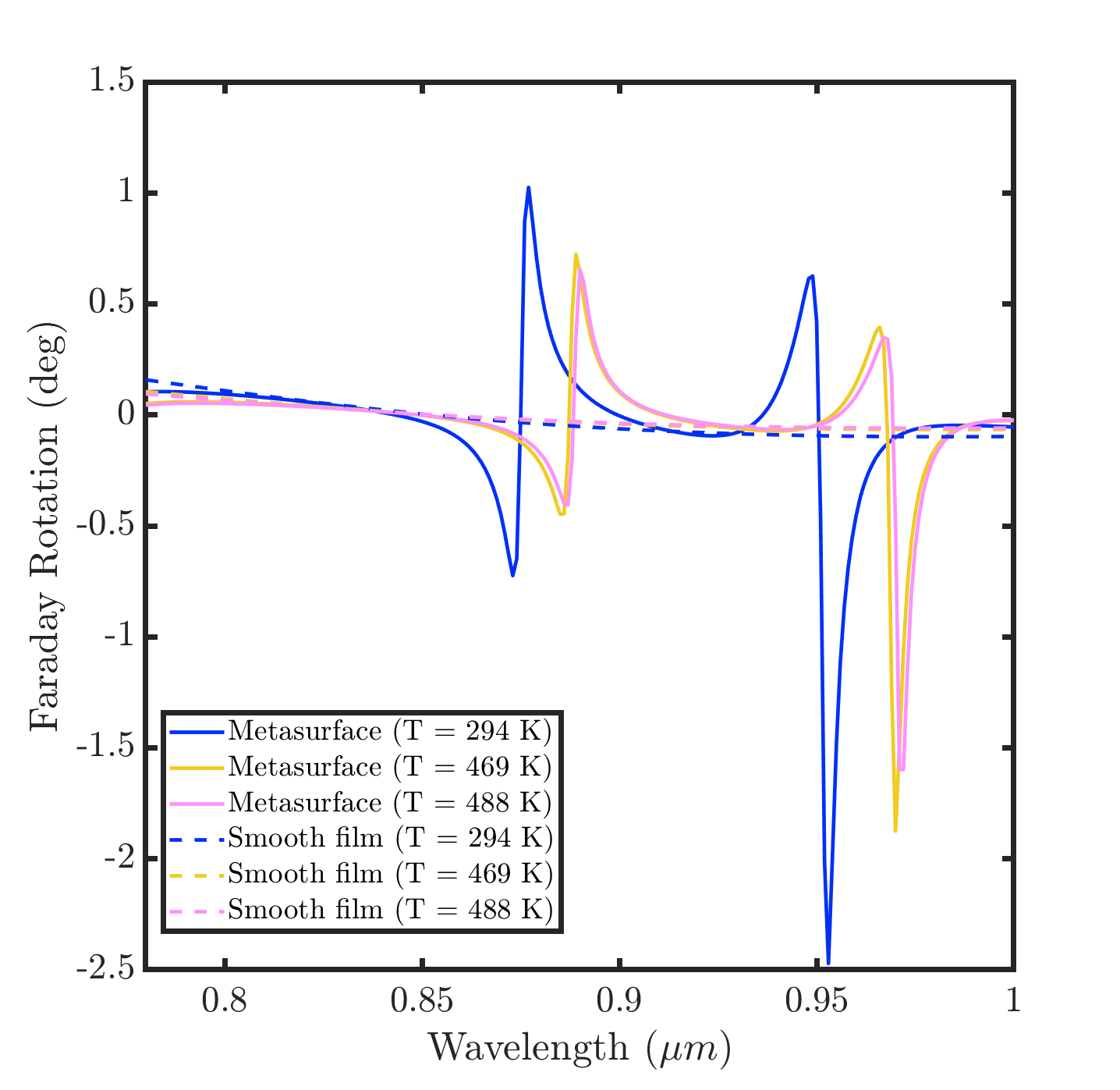}\\
(c)~~~~~~~~~~~~~~~~~~~~~~~~~~~~~~~~~~~~~~~~~~~~~~~~~~~~~~~~~~~~~~~~~~~~~~~~~(d)\\
     \includegraphics[width=0.45\linewidth]{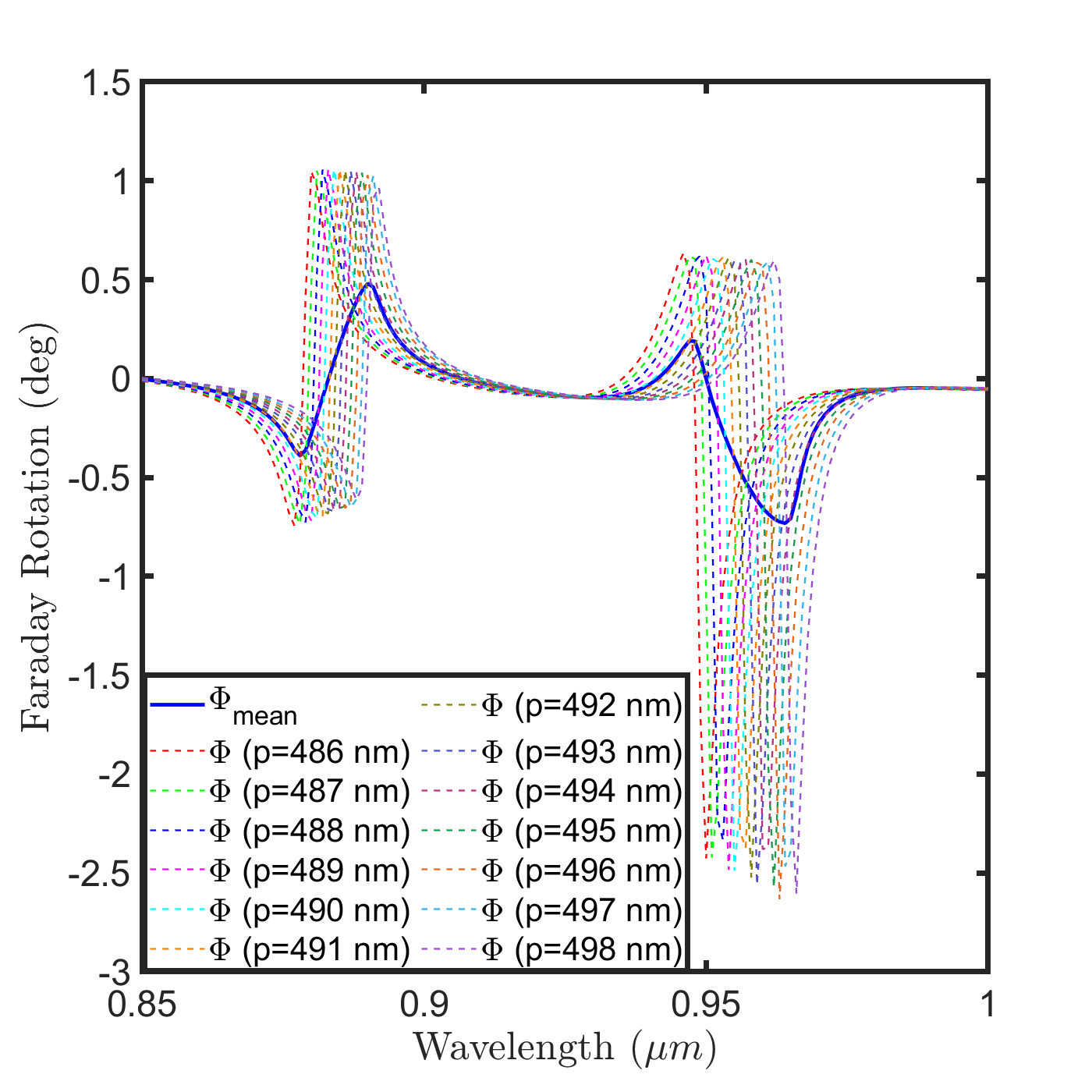}
          \includegraphics[width=0.45\linewidth]{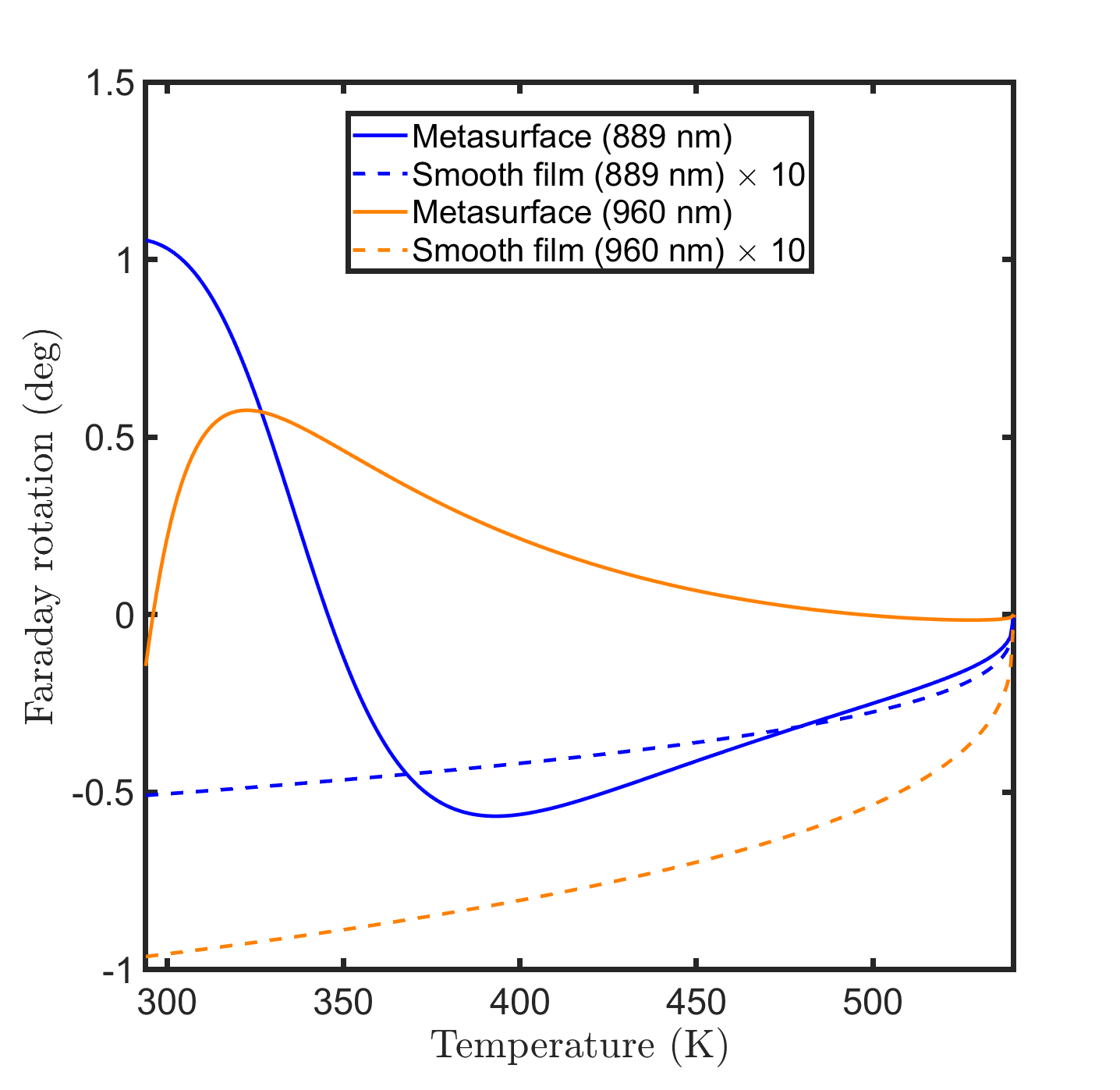}

    \caption{(a) Experimental spectra of Faraday rotation in metasurface measured at different temperatures: 294 K, 469 K, and 488 K. The dashed lines represent the Faraday rotation of a smooth film measured at the same temperatures (b) Numerical spectra of Faraday effect simulated using the RCWA method for the same temperatures. (c)  Spectra of Faraday rotation in the metasurface for the periods $P=492\pm5$, including the averaged value.(d) Numerically simulated temperature dependence of the Faraday rotation for 889 nm and 960 nm wavelengths.}
    \label{Fig: Faraday}
\end{figure*}

Mie resonances are observed as the wide angle-independent dips in the transmittance spectra (see Fig.\ref{Fig: Modes in metasurface}c,d) at 875 nm and 940 nm (952 nm in simulations), correspondingly. The resonance at 940 nm corresponds to the excitation of the ED mode (see Fig.\ref{Fig: Modes in metasurface}f), while MD mode is excited at 875 nm (see Fig.\ref{Fig: Modes in metasurface}e). 

Due to the peculiar character of the displacement currents induced in the Dy:CeYIG film by the optical fields under optical mode excitation~\cite{xia2022enhancement}, Faraday rotation is significantly enhanced in the vicinity of the Mie-type resonances. The Faraday rotation spectra (Fig.~\ref{Fig: Faraday}a) has a typical for a metasurface Fano-type resonance due to the presence of the non-resonant contribution of the smooth film and the resonant contribution arising from the excitation of the optical modes.

The unusual feature of the considered all-dielectic metasurface is that its Faraday rotation spectra has a sharp resonance with a high derivative of 0.02~deg/nm in the vicinity of the resonance. For a comparison, Faraday rotation of the smooth film changes smoothly with 0.002~deg/nm derivative in the same range. The position of the magneto-optical resonance is determined by the metasurface parameters providing the excitation of the corresponding mode. Therefore, the resonance position could be efficiently controlled using the external stimulus, for example, thermal heating.

Heating changes the magnetization of ferrimagnetic Dy:CeYIG, and its gyration changes consequently. We estimate the Curie temperature as $T_\mathrm{C}=540K$, and thus consider the gyration to vary with temperature as $g_\mathrm{Dy:CeYIG}=g_\mathrm{Dy:CeYIG}^{0} \cdot (1-(T/T_\mathrm{C}))^{0.33}$ ~\cite{stohr2006magnetism}. In the smooth films such a dependence leads to the proportional decrease of the Faraday rotation angles with temperature. However, the situation changes in the presence of a metasurface.

The metasurface is sensitive to the heating process that simultaneously changes the permittivity of the materials and the periodicity due to the thermal expansion of the media, resulting in the shift of the resonance position~\cite{zograf2021all}. In contrast to the transmittance Mie-type resonances, which are quite broad and thus poorly affected by the refractive index changes, the Faraday rotation resonances are sharp and sensitive to heating. The shift of the Mie mode position in optical spectra due to the heating process leads to the shift of a sharp Fano resonance Faraday rotation curve. This shift equal to $\sim0.04$nm/K is clearly seen in experimentally measured spectra shown in Fig.~\ref{Fig: Faraday}a and also verified by the numerical simulations shown in Fig.~\ref{Fig: Faraday}b. As various fabrication inaccuracies are not taken into account in sumulations, the quality factor of the Faraday effect resonances and the enhancement of the Faraday rotation in theory is higher that observed experimentally. At the same time, the resonance position and its thermal shift are qualitatevely the same. Fig.~\ref{Fig: Faraday}c shows that averaging over quite moderate spread of the periods $P=492\pm5$~nm mimicking fabrication inaccuracies is taken into, the magneto-optical resonance becomes broader and its Q-factor and magnitude becomes close to what is observed experimentally. Fig.~\ref{Fig: Faraday}c shows that eliminating various fabrication imperfections might result in a very sensitive temperature dependence of the Faraday rotation with a derivative up to $\partial \Phi / \partial T \approx 0.03~\mathrm{deg/K}$.

A significant difference in the Faraday rotation spectra measured at different temperatures is observed in the vicinity of the MD resonance at 950 nm. The most interesting range is the wavelengths 940-950 nm, where $\Phi(T=294K)<0$ and $\Phi(T=488K)>0$ and the difference between these rotation angles reaches 0.35~deg at 945~nm wavelength. This difference is higher than, for example, the Faraday rotation in the smooth film $|\Phi|=0.07^\circ$ for the same wavelength. This is quite an unusual result that changing the temperature of metasurface with constant-sign gyration one can get different signs of the Faraday effect for the same wavelength of incident light. For example, changing the temperature from the room to the 488K results in the gradual changes of Faraday rotation values from -0.297~deg and +0.056~deg for fixed wavelength of 945 nm observed experimentally. Such a continuous change of the Faraday rotation is verified in simulations (Fig.~\ref{Fig: Faraday}d). Therefore, not only the value, but also the sign of the Faraday rotation in the considered metasurface is efficiently controlled by its temperature. 

An inversion of the sign of the Faraday effect can be analyzed in the framework of the coupled mode theory (CMT), similar to the way the mechanism of the inversion of the transverse Kerr effect was interpreted earlier~\cite{tang2024giant,darki2016temporal}. According to the Mie theory, electromagnetic fields inside the structure are expressed in terms of dimensionless values of $k_0nr$, where $k_0$ is the wavenumber in vacuum, $n$ is the refractive index of the medium, and $r$ is the characteristic size of the system. As the refractive index in the magnetic film for the two circular components is different ($n_{\pm}=n_\mathrm{Dy:CeYIG}\pm\frac{g_\mathrm{Dy:CeYIG}}{2n_\mathrm{Dy:CeYIG}}$~\cite{zvezdin1997modern}), Mie resonance frequencies also slightly differ. This leads to a characteristic S-shaped phase spectral response, and, consequently, S-shaped Faraday rotation spectra as $\Phi=\frac{1}{2}\arg(\frac{t_+}{t_-})$, where $t_{\pm}$ is amplitude transmission coefficient. Due to the different resonant wavelengths, phase responses of these components differ, which leads to the presence of a special point at which their difference turns to zero, and, consequently, Faraday rotation also turns to zero. When the parameters of the resonance interaction are further changed (e.g., by varying the temperature), the phase difference at this wavelength changes sign, leading to a change in the sign of the Faraday effect.

\section{Discussion and Applications}

\subsection{All-optical control of the Faraday rotation}

The temperature-controlled Faraday rotation described above can be used for the advanced light modulation devices when the Faraday polarization rotation of the signal is controlled by heating via either an external laser or by the intensity of the signal itself. 

Metasurface heating by $\Delta T$ temperature by a laser pulse with $W_\mathrm{light}$ enegry is described as:
\begin{equation}
    \Delta T=\frac{ A \cdot W_\mathrm{light} }{ (m_{\text{Dy:CeYIG}}\cdot c_{\text{Dy:CeYIG}}+m_{\text{Si}}\cdot c_{\text{Si}})},
\end{equation}
where $A$ is an absorption coefficient, $m_{\text{Dy:CeYIG}}, m_{\text{Si}}$ are masses of CeYIG layer and silicon nanocylinders, respectively,  $c_{\text{Dy:CeYIG}}, c_{\text{Si}}$ are their specific heat capacities (see Appendix for the exact values). We considered the diameter of the beam equal to 10 microns, and the pulse repetition rate equal to 80 MHz. Substituting the parameters of the structure and heat capacity of the materials showed that to heat by 194 K, it is necessary to transfer energy  $A \cdot W_\mathrm{light}\approx 12$nJ which is acessible for various types of lasers. Let us consider the two most interesting cases, when this power is transferred by an external control laser with the wavelength corresponding to the absorption band of the structure, or by the signal laser itself.

\begin{figure}[htb]
\centering

\includegraphics[width=0.9\linewidth]{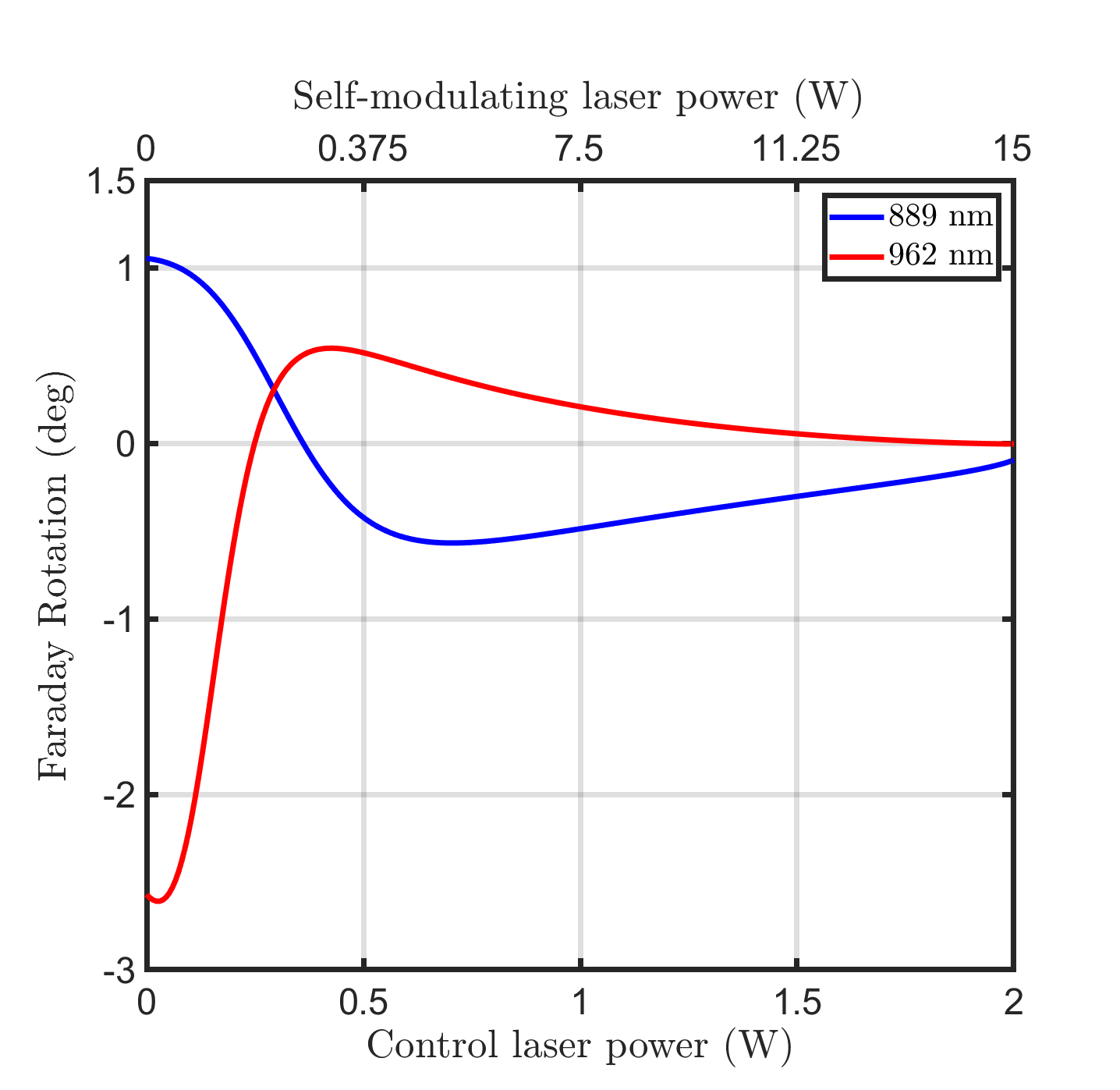}
\caption{Faraday rotation angle power dependence for the cases of (bottom axis) a control 532-nm laser heating the structure for the signal laser of 889 nm and 962 nm wavelength (see the legend); (upper axis) self-modulation of signal laser having 889 nm and 962 nm, correspondingly.}
\label{fig:P_950_532}
\end{figure}

One of the ways to heat the sample is to use a control laser with a wavelength inside the absorption band, for example, 532 nm. Due to the presence of silica cylinders and the nanopattern, the absorption at this wavelength is as high as $A \sim 70\%$. Figure~\ref{fig:P_950_532} (bottom axis) shows the dependence of Faraday rotation on the power of the 532 nm control laser, which heats the sample.

At low control laser powers, the Faraday rotation is negative $(\Phi(P<P_0)<0)$, while increasing the power makes the Faraday rotation angle positive $(\Phi(P>P_0)>0)$. The threshold power $P_0$, at which the Faraday rotation changes sign, is $P_0 = 0.37$~W and $P_0 = 0.25$~W for signal lasers with 962~nm and 889~nm wavelengths, correspondingly.

Such an approach also gives rise to a novel type of tunable magneto-optical metasurfaces whose magneto-optical response is not fixed after the structure fabrication but can be tuned in a wide range via the external stimulus. It is important that thermal heating can be performed locally at the spots down to $\sim1~\mu$m diameter using focused laser pulses. In this case, the magneto-optical response can be tuned locally by the proper selection of the local temperature in the area of several microns. Thus one may obtain the inhomogeneous Faraday rotation distributions across the metasurface by the creation of the laser beam patterns with the desired intensity profiles.

Another possibility is the self-modulation of the Faraday rotation by a signal pulse. Figure~\ref{fig:P_950_532} (upper axis) shows self-modulating Faraday polarization rotation effects for the case if the structure is heated by the signal laser itself having the wavelength 889~nm and 962~nm, correspondingly. Qualitatevely, the Faraday rotation dependence on the power of the laser pulse heating the metasurface is similar for the cases of the presence of an additional control laser or self-modulation effects. However, as the absorption in infrared range is smaller, $A\sim 16-20\%$ for the resonant wavelengths around 889~nm and 962~nm, the threshold power $P_0$ required for observation of the Faraday rotation sign change using self-modulation effect is 6-7 times higher, $P_0=2.7$~W and $P_0=1.6$~W, correspondingly.

For the laser beams with wavelengths of 889 nm and 962 nm, different signs and values of the Faraday effect can be obtained depending on the power of the laser illuminating the metasurface. Thus, quite interesting situation is observed where the low-power pulses experience negative-sign Faraday rotation, and high-power pulses experience positive-sign Faraday rotation. Adding an additional polarizer after such a metasurface allows to control the output intensity of the pulses.

\subsection{Sensing via Faraday rotation in metasurface}

High-Q magneto-optical resonances are always attracting for sensing devices~\cite{rizal2021magnetophotonics, tang2023enhancing, borovkova2019high, merzlikin2018magneto, danilin2021magneto, ignatyeva2020all, ignatyeva2022all}. Faraday rotation spectra exhibits very high derivative in the vicinity of the MD resonance, and its position and magnitude are very sensitive to the parameters describing the system itself (temperature, magnetic field) and its environment. This enables construction of various types of sensors on the base of the designed metasurface.

 \begin{figure*}[htb]
 \centering
(a)~~~~~~~~~~~~~~~~~~~~~~~~~~~~~~~~~~~~~~~~~~~~~~~~~~~~~~~~~~~~~~~~~~~~~~~~~~~~~~~~~~~~~~~~~~~~~~~~~~~~~~~~~~~~~~(b)\\   
\includegraphics[width=0.45\linewidth]{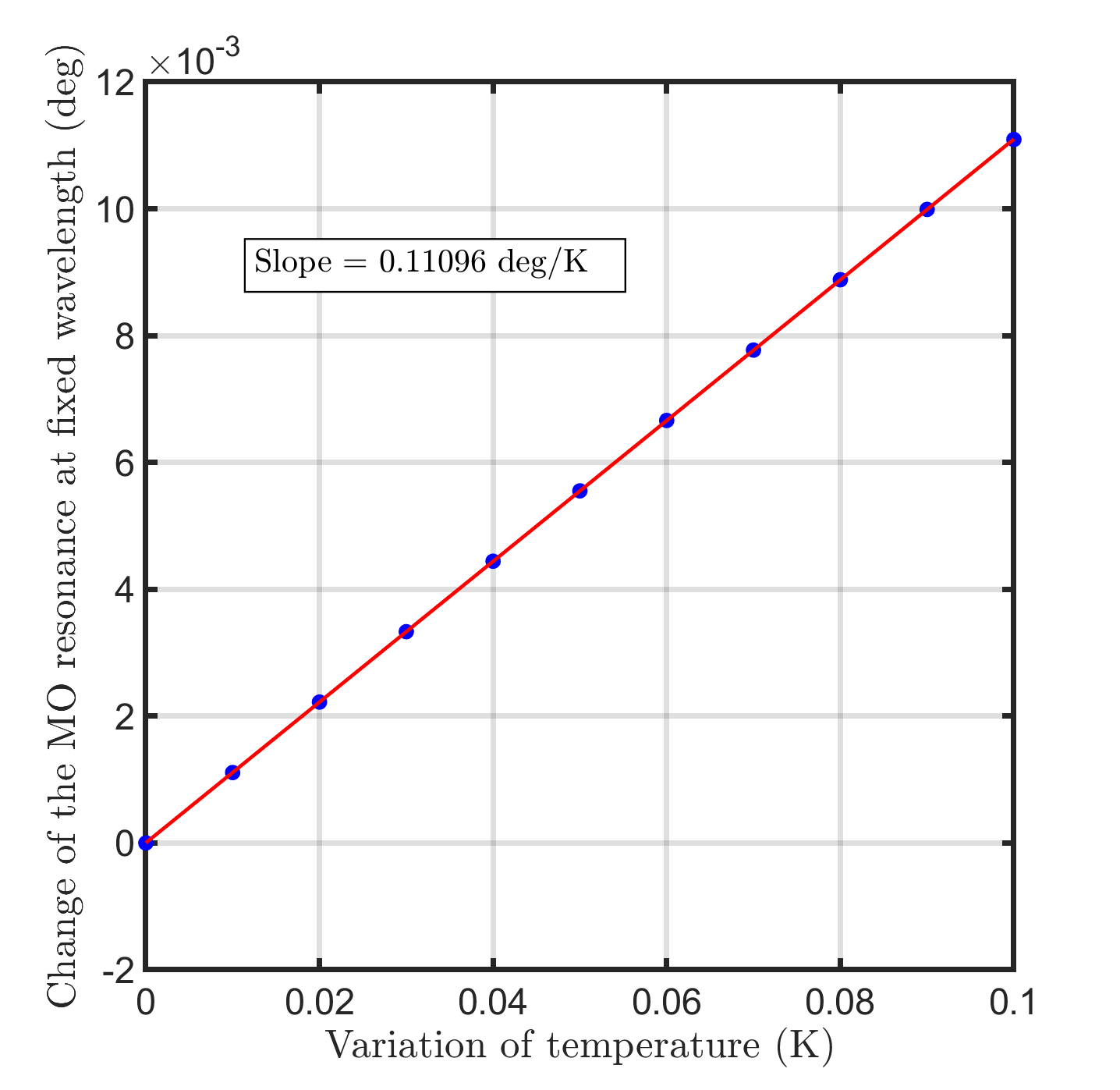}
\includegraphics[width=0.45\linewidth]{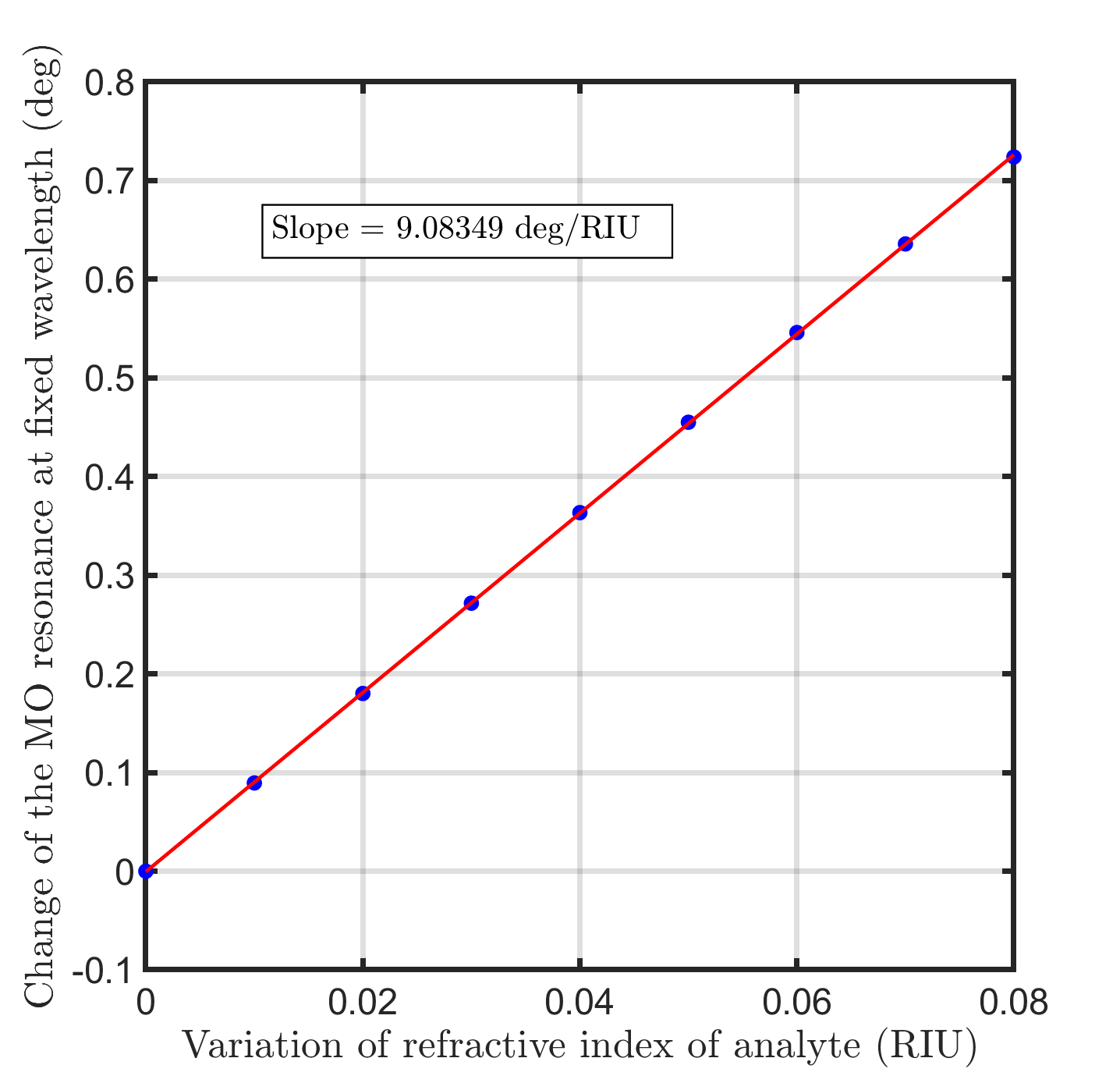}
\caption{Temperature (a) and refractive index (b) sensing}
\label{Fig: Sensors}
\end{figure*}

{\it Temperature sensor.} The most straightforward application of such a metasurface is a temperature sensor. For a fixed wavelength, the Faraday effect in metasurface changes with temperature so that the sensitivity of the such a temperature sensor is approximately $\partial \Phi / \partial T =0.111 \;\mathrm{deg/K}$ at $\lambda=0.9565$~nm wavelength (see Fig.~\ref{Fig: Sensors}a). The advantage of such a magnetophotonic-based sensor is an ability to measure the temperature distribution with a very high spatial resolution up to several microns and high precision. Nowadays, Faraday rotation angles down to $\sim10^{-5}$~deg can be measured, which means that small temperature changes $\Delta T \sim 10^{-4}$~K can be detected using the presented metasurface.

{\it Refractive index sensor.} Although Mie-type modes are concentrated predominantly inside the nanocylinders, laser radiation also penetrates in the gaps between them (see Fig.~\ref{Fig: Modes in metasurface}e,f), which makes the structure sensitive to the refractive index of the environment. The resonance shifts under the variation of the environmental refractive index as $\partial \lambda_\mathrm{MD}/ \partial n_\mathrm{env} =5.7~\mu\mathrm{m/RIU}$. Thus, refractive index might be determined by the measurement of the magneto-optical response at a fixed wavelength, which varies linearly with the refractive index of the environment (see Fig.~\ref{Fig: Sensors}b). The sensitivity of such a sensor is $\partial \Phi / \partial n_\mathrm{env} =9.1$~deg/RIU reveals the sensor's high sensitivity to refractive index fluctuations. 
One of the important sensors characteristics is FOM (Figure of Merit) defined as the ratio of sensitivity to the width of the resonant peak $     \mathrm{FOM}=(\partial \lambda_\mathrm{MD} / \partial n_\mathrm{env}) (\Delta \lambda_\mathrm{MD})^{-1}$. The proposed metasurface-based sensor exhibits $\mathrm{FOM}=191.92\;RIU^{-1}$ in the experiment and  $\mathrm{FOM}=1140\;RIU^{-1}$ in the simulations. The threshold variations of the refractive index which can be measured via the sensor are $\Delta n \sim \times 10^{-6}$~RIU. The values of the sensitivity and FOM are comparable with the other magneto-optical devices~\cite{rizal2021magnetophotonics}, however, further tuning of the structure design and reducing of the fabrication inaccuracies might increase the performance. In contrast to the well-known plasmonic sensors, operating frequencies of such a hybrid Si+Dy:CeYIG lie in the infrared range, which might be important for the analysis of the fingerprints of resonances of various organic molecules~\cite{ falek2021chip}.

\section{Conclusion}

A hybrid metasurface containing a smooth Dy:CeYIG layer and a Si nanodisk pattern on its top provides sharp Faraday effect Fano-type resonances, which positions are extremely sensitive to the metasurface parameters. Here we demonstrate the metasurface magneto-optical response tunability via temperature changes. We show both experimentally and numerically that not only the value, but also the sign of the Faraday rotation of the metasurface could be efficiently tuned between -0.3~deg and +0.1~deg values by the variation of the metasurface temperature. 

Possible applications of such a temperature-sensitive metasurface are shown, such as an ability to perform modulation and self-modulation of the polarization rotation by the intensity of the laser radiation. An ability to perform temperature and refractive index sensing based on the proposed metasurface is also demonstrated.

\textbf{Author Contributions:} Conceptualization; formal analysis; investigation; methodology; project administration; supervision; writing—original draft; writing—review and editing. All authors have read and agreed to the published version of the manuscript.\\
\textbf{Funding:} This research was funded by the Ministry of Science and Higher Education of Russian Federation, project N 23-12-00310.\\
\textbf{Institutional Review Board Statement:} Not applicable.\\
\textbf{Informed Consent Statement:} Not applicable.\\
\textbf{Data Availability Statement:} The data presented in this study are available on request from the corresponding author.\\
\textbf{Acknowledgments:} The authors thank A.N. Kalish and D.M. Krichevsky for the fruitful discussions and assistance with the modeling.\\
\textbf{Conflicts of Interest:} The authors declare no conflict of interest.

\newpage
\section*{Appendix A. Optical and magneto-optical Properties of Dy:CeYIG and Si}

The following optical and magneto-optical properties of the Dy:CeYIG were obtained from the experimental measurements of the smooth films and used in the simulations.

 \begin{figure}[htb]
 ~~~~~~~~~~~~~~~~~~~~~~~~~~~~~~~~~~~~~(a)~~~~~~~~~~~~~~~~~~~~~~~~~~~~~~~~~~~~~~~~~~~~~~~~~~~~~~~~~~~~~~~~~~~~~~~~~(b)\\
 \includegraphics[width=0.49\linewidth]{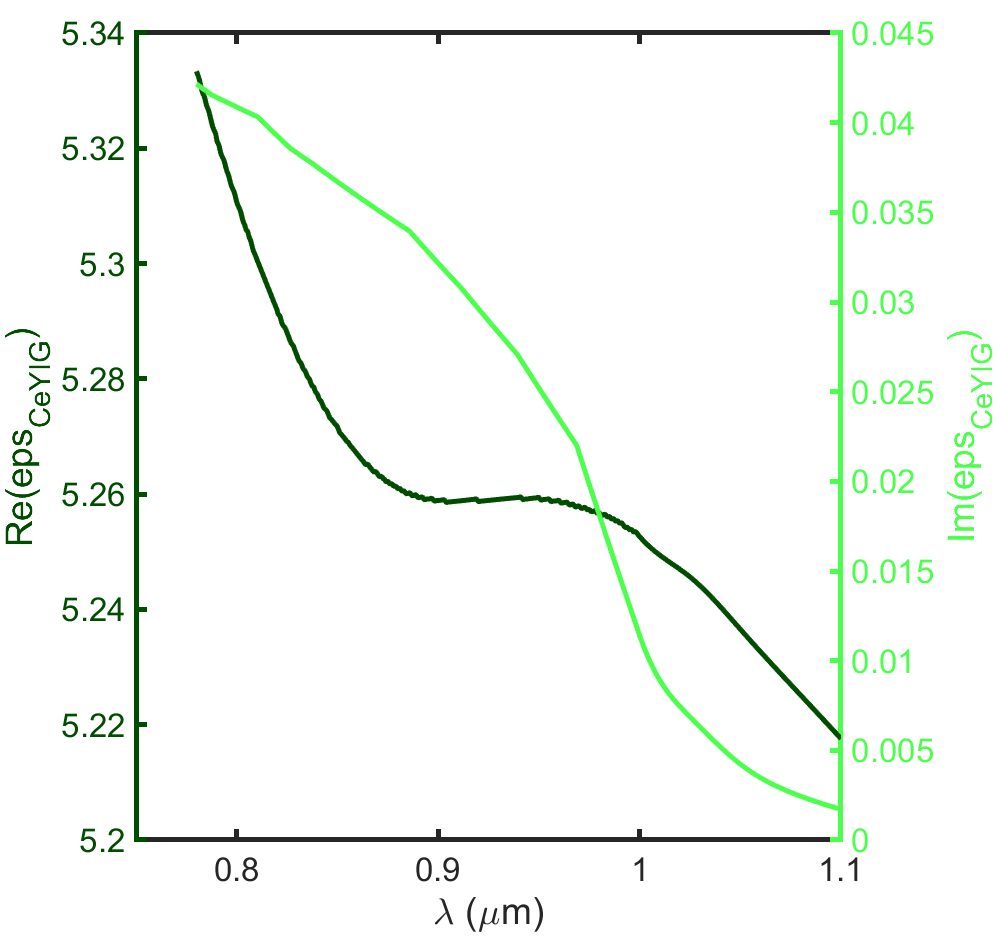}
 \includegraphics[width=0.49\linewidth]{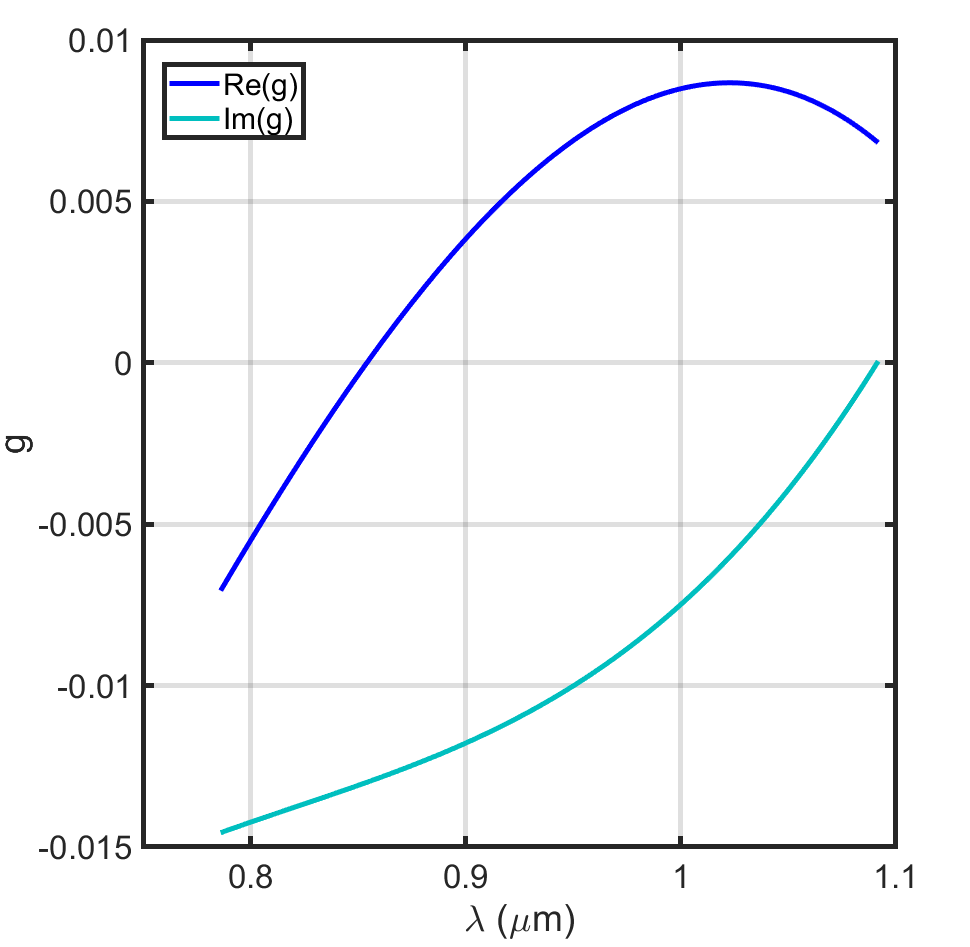}
    \caption{Dielectric permittivity (a) and gyration (b) of Dy:CeYIG}
    \label{fig:eps_ge}
\end{figure}

\begin{figure}[htb]
~~~~~~~~~~~~~~~~~~~~~~~~~~~~~~~~~~~~~(a)~~~~~~~~~~~~~~~~~~~~~~~~~~~~~~~~~~~~~~~~~~~~~~~~~~~~~~~~~~~~~~~~~~~~~~~~~(b)\\
\includegraphics[width=0.49\linewidth]{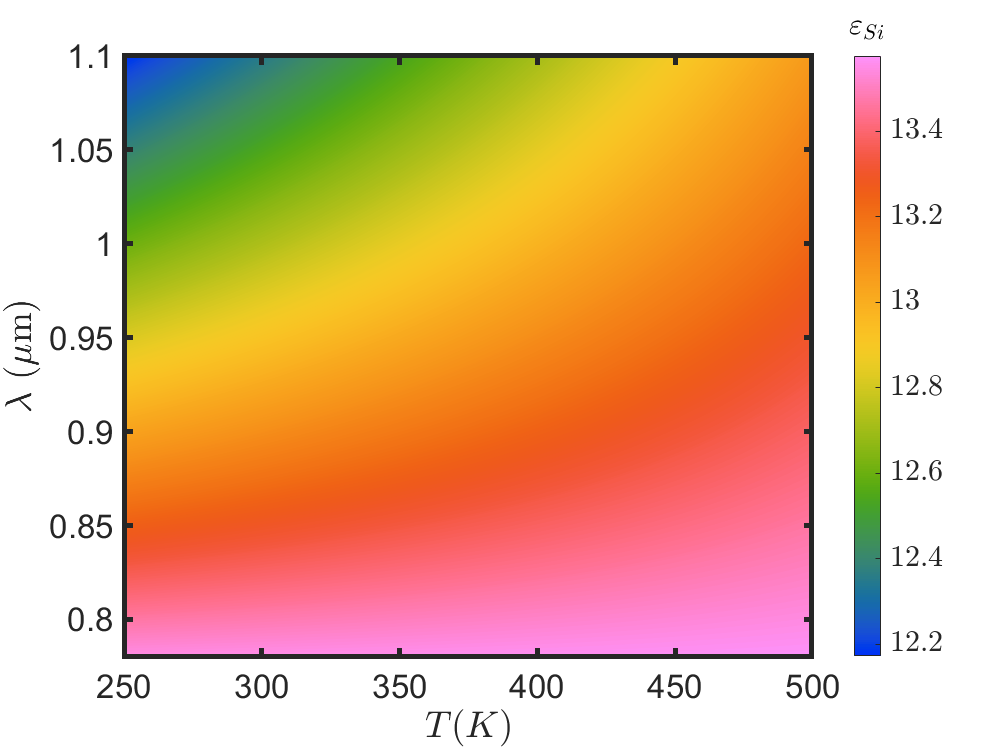}
 \includegraphics[width=0.49\linewidth]{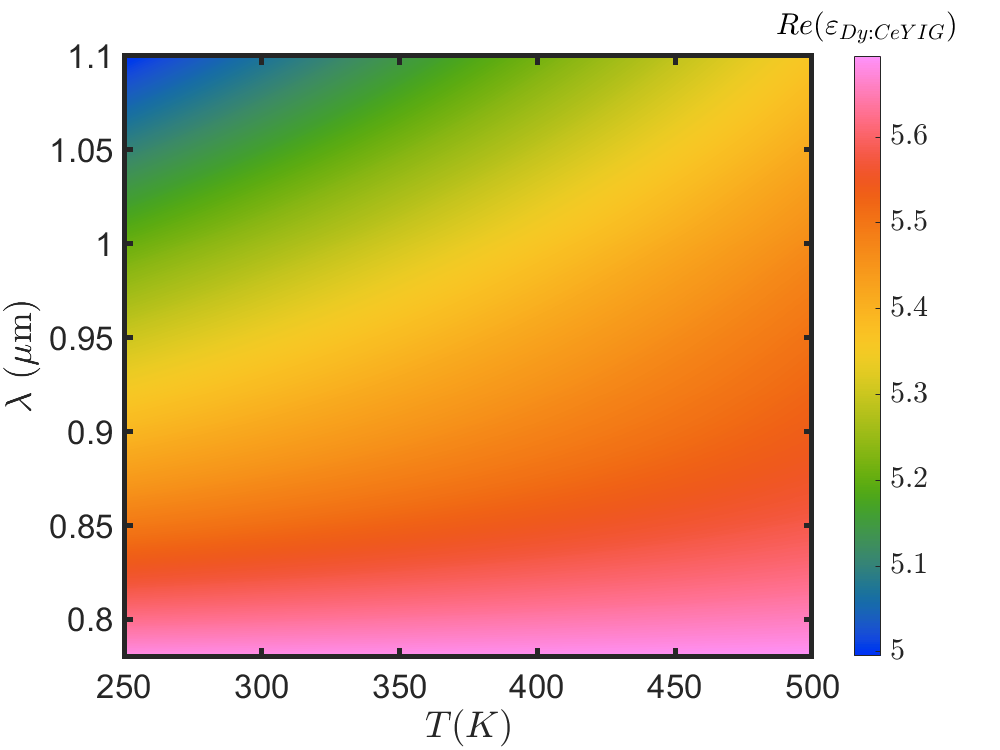}
    \caption{Dielectric permittivity spectra of Si (a) and Dy:CeYIG (b)}
    \label{fig:eps_Si}
\end{figure}

The temperature dependence of the permittivity of Dy:CeYIG and Si is taken into account~\cite{franta2017temperature} as shown in Fig.~\ref{fig:eps_Si}.

\section*{Appendix B. Numerical simulations}

\begin{figure}[htb]
  \centering
 \includegraphics[width=0.99\linewidth]{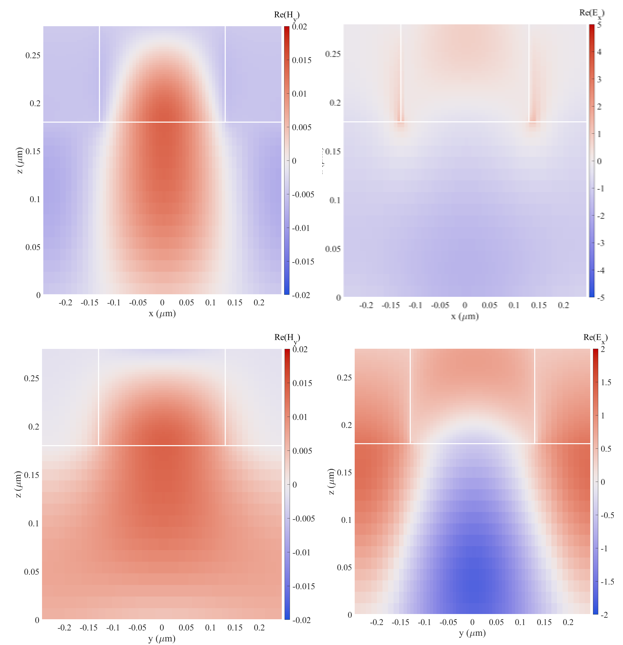}
       \caption{The real components of H$_y$ and E$_x$ fields are calculated for a structure cross-section drawn through the center of a cylinder}
    \label{fig:Hy- and Ex-field}
\end{figure}

\begin{figure}[htb]
 ~~~~~~~~~~~~~~~~~~~~~~~~~~~~~~~~~~~~~(a)~~~~~~~~~~~~~~~~~~~~~~~~~~~~~~~~~~~~~~~~~~~~~~~~~~~~~~~~~~~~~~~~~~~~~~~~~(b)\\
 \includegraphics[width=0.49\linewidth]{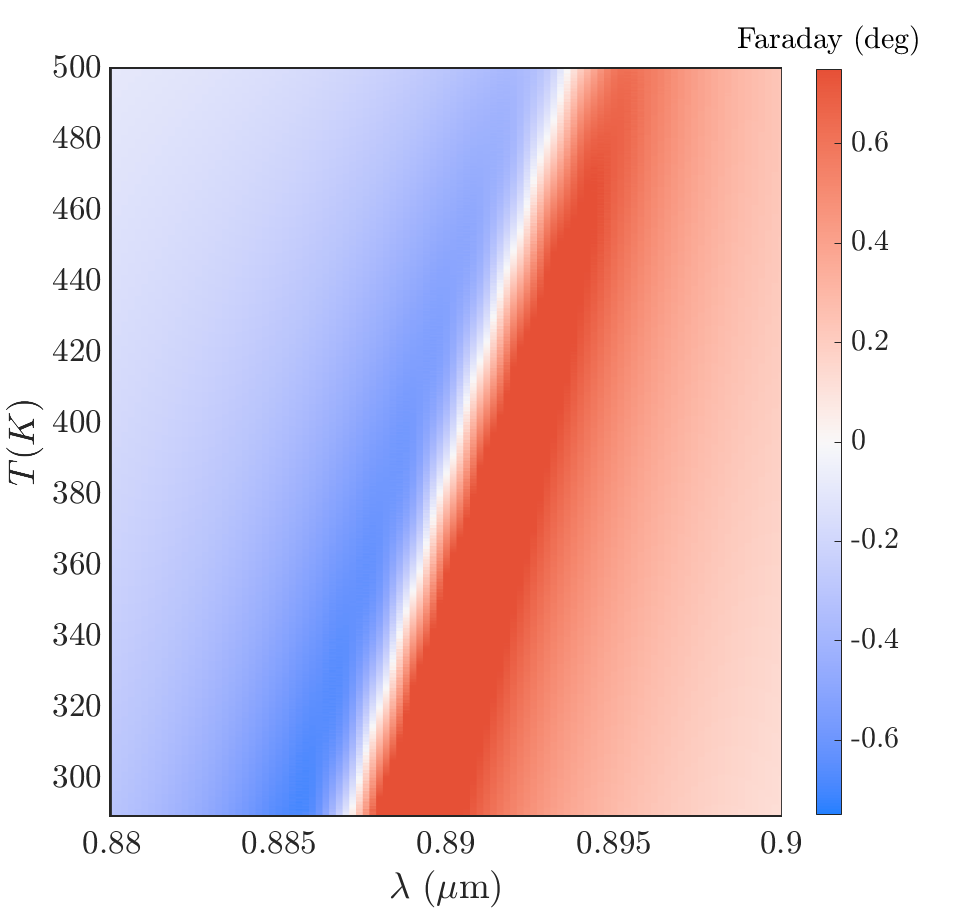}
    \includegraphics[width=0.49\linewidth]{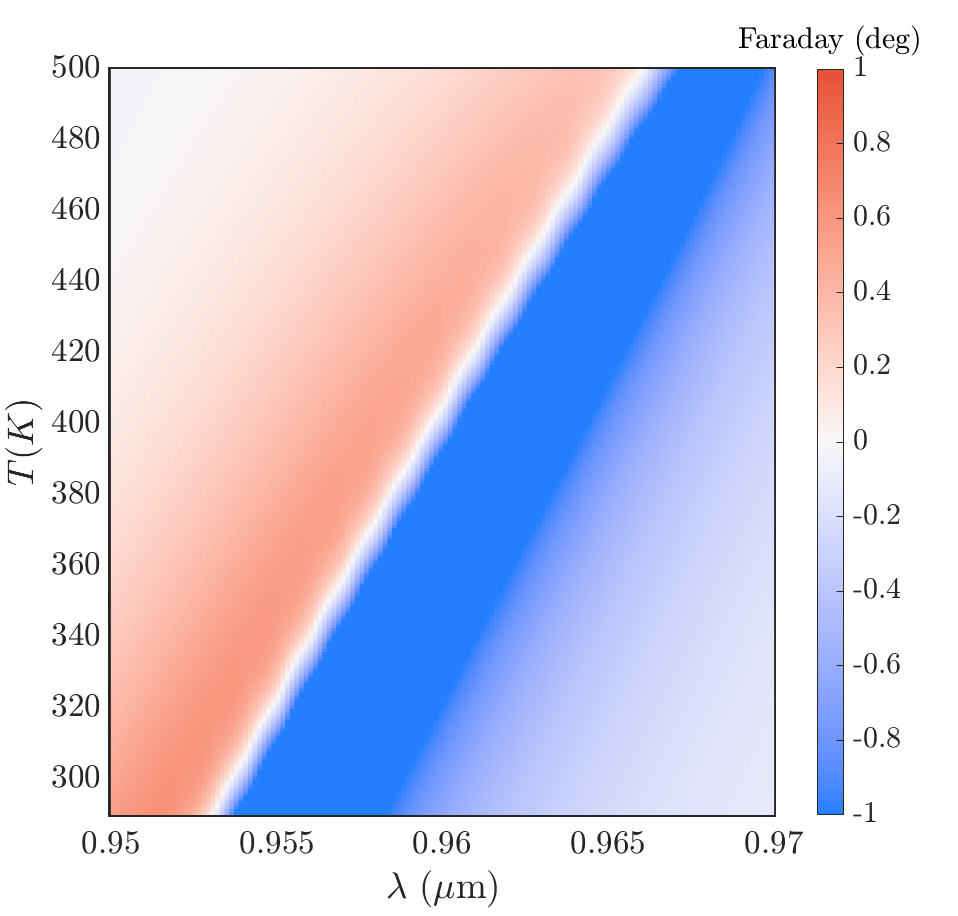}
    \caption{Numerical Faraday rotation spectra for a two-dimensional nanophotonic structure at various wavelengths. (a) Spectral response near the first resonant peak, encompassing the wavelength range of 0.88 to 0.9 $\mu m$; (b) Spectral response near the second resonant peak, spanning wavelengths from 0.95 to 0.97 $\mu m$.}
    \label{fig:2peak}
\end{figure}

The results of numerical simulations were obtained using the Rigorous Coupled-Wave Analysis (RCWA) method, which enabled electromagnetic numerical simulations of the structure's optical and magneto-optical properties. The modeling parameters have been adjusted as follows: the Si disk is characterized by a height of 100 nm and a radius of 130 nm. To simplify the model, the Dy:CeYIG and YIG layers were modeled as a unified magnetic layer, 180 nm in thickness, with its dielectric permittivity tensor detailed in Appendix A. The dielectric permittivity of the glass substrate was set at 2.1025. Some of the additional numerical results are shown in Figs.~\ref{fig:Hy- and Ex-field} and \ref{fig:2peak}.

\section*{Appendix C. Thermal Characteristics of CeYIG and Silicon nanocylinders}

In the thermal analysis presented in the main text, the mass and specific heat capacity of the CeYIG layer and silicon nanocylinders are key parameters influencing the heat dynamics of the system. The mass of each material is determined based on its volume and density. The exact values of these parameters are summarized in Table~\ref{tab:thermal_properties}.  

\begin{figure*}[htb]
\centering

\includegraphics[width=0.7\linewidth]{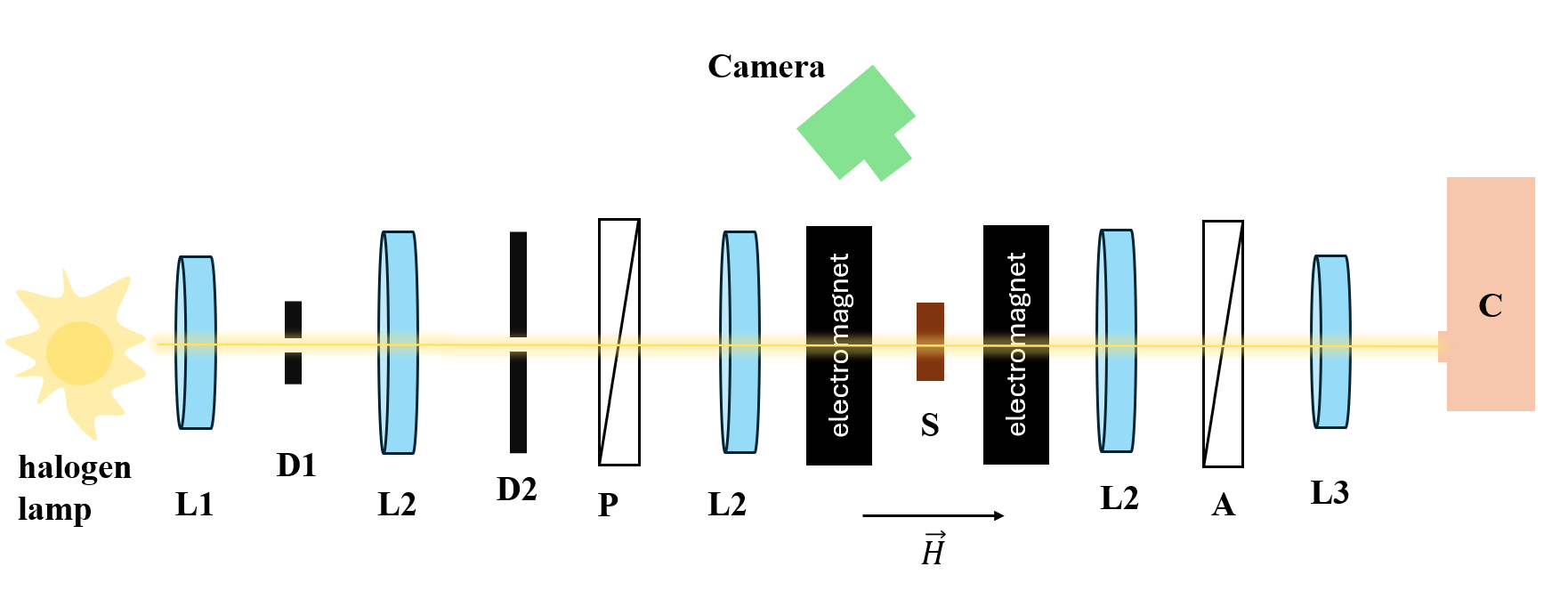}
\caption{Experimental setup}
\label{fig:experiment}
\end{figure*}

The mass of the CeYIG layer \( m_{\text{CeYIG}} \) is calculated as:  
\begin{equation*}
    m_{\text{CeYIG}} = V_{\text{CeYIG}} \cdot \rho_{\text{CeYIG}}
\end{equation*}
where \( V_{\text{CeYIG}} \) is the volume of the layer, determined by its thickness and lateral dimensions, and \( \rho_{\text{CeYIG}} = 7.1 \) g/cm$^3$ is the material density. The computed mass value is:  
\begin{equation}
    m_{\text{CeYIG}} = 1.004 \times 10^{-13} \text{ kg}
\end{equation}  

Similarly, the mass of the silicon nanocylinders \( m_{\text{Si}} \) is calculated as:  
\begin{equation}
    m_{\text{Si}} = V_{\text{Si}} \cdot \rho_{\text{Si}}
\end{equation}
where \( V_{\text{Si}} \) is the total volume of all nanocylinders, considering their cylindrical geometry, and \( \rho_{\text{Si}} = 2.33 \) g/cm$^3$ is the silicon density. The computed mass value is:  
\begin{equation}
    m_{\text{Si}} = 4.008 \times 10^{-15} \text{ kg}
\end{equation}

The specific heat capacities of CeYIG and silicon at room temperature are assumed to be \( c_{\text{CeYIG}} = 600 \) J/(kg·K) and \( c_{\text{Si}} = 800 \) J/(kg·K), respectively.  

A summary of the thermophysical parameters is provided in Table~\ref{tab:thermal_properties}.  

\begin{table}[h]

    \caption{Thermophysical properties of CeYIG and silicon nanocylinders.}
    \label{tab:thermal_properties}
    \begin{tabular}{|l|c|c|}
        \hline
        \textbf{Parameter} & \textbf{Symbol} & \textbf{Value} \\
        \hline
        Density of CeYIG & \( \rho_{\text{CeYIG}} \) & 7.1 g/cm³  \\
        Density of Si & \( \rho_{\text{Si}} \) & 2.33  g/cm³ \\
        Specific heat capacity of CeYIG & \( c_{\text{CeYIG}} \) & 600 J/(kg·K) \\
        Specific heat capacity of Si & \( c_{\text{Si}} \) & 800 J/(kg·K) \\
        Mass of CeYIG layer & \( m_{\text{CeYIG}} \) & \( 1.004 \times 10^{-13} \) kg \\
        Mass of silicon nanocylinders & \( m_{\text{Si}} \) & \( 4.008 \times 10^{-15} \) kg \\
        \hline
    \end{tabular}
\end{table}

\section*{Appendix D. Experimental measurements}

The experiment was conducted to investigate the Faraday effect in a transmission configuration using a halogen lamp, an optical system for focusing and polarization, and a detection system to measure changes in the intensity of transmitted radiation. At the system's input, a halogen lamp OceanOptics HL 2000-HP was used as the light source. The beam is formed using a set of lenses (L1, L2, L3) and apertures (D1, D2). The polarization properties of the sample are examined using the Thorlabs LPVIS050 polarizer (P) and an analyzer (A). The light is then focused with a lens having a focal length of 100 mm and is subsequently recorded using a spectrometer (C) with a range of 200–1100 nm.

The sample (S) was placed in a special copper holder, ensuring temperature stabilization. A heating element is integrated into the holder. Temperature control is performed using a platinum temperature sensor Pt-1000, connected to a temperature controller. The temperature controller stabilizes the set temperature with high accuracy. The transmitted radiation is measured using a spectrometer with a CCD camera. Experiment parameters (wavelength adjustment, temperature control, polarization adjustment) were set and controlled by an automated system connected to a computer. Data were recorded and analyzed using specialized software.

This configuration of the experimental setup allowed for highly accurate measurement of changes in the intensity and polarization of light passing through the sample, as well as precise control of the temperature conditions during the experiment.
\newpage
\bibliography{apssamp}

\begin{thebibliography}{41}%
\makeatletter
\providecommand \@ifxundefined [1]{%
 \@ifx{#1\undefined}
}%
\providecommand \@ifnum [1]{%
 \ifnum #1\expandafter \@firstoftwo
 \else \expandafter \@secondoftwo
 \fi
}%
\providecommand \@ifx [1]{%
 \ifx #1\expandafter \@firstoftwo
 \else \expandafter \@secondoftwo
 \fi
}%
\providecommand \natexlab [1]{#1}%
\providecommand \enquote  [1]{``#1''}%
\providecommand \bibnamefont  [1]{#1}%
\providecommand \bibfnamefont [1]{#1}%
\providecommand \citenamefont [1]{#1}%
\providecommand \href@noop [0]{\@secondoftwo}%
\providecommand \href [0]{\begingroup \@sanitize@url \@href}%
\providecommand \@href[1]{\@@startlink{#1}\@@href}%
\providecommand \@@href[1]{\endgroup#1\@@endlink}%
\providecommand \@sanitize@url [0]{\catcode `\\12\catcode `\$12\catcode `\&12\catcode `\#12\catcode `\^12\catcode `\_12\catcode `\%12\relax}%
\providecommand \@@startlink[1]{}%
\providecommand \@@endlink[0]{}%
\providecommand \url  [0]{\begingroup\@sanitize@url \@url }%
\providecommand \@url [1]{\endgroup\@href {#1}{\urlprefix }}%
\providecommand \urlprefix  [0]{URL }%
\providecommand \Eprint [0]{\href }%
\providecommand \doibase [0]{https://doi.org/}%
\providecommand \selectlanguage [0]{\@gobble}%
\providecommand \bibinfo  [0]{\@secondoftwo}%
\providecommand \bibfield  [0]{\@secondoftwo}%
\providecommand \translation [1]{[#1]}%
\providecommand \BibitemOpen [0]{}%
\providecommand \bibitemStop [0]{}%
\providecommand \bibitemNoStop [0]{.\EOS\space}%
\providecommand \EOS [0]{\spacefactor3000\relax}%
\providecommand \BibitemShut  [1]{\csname bibitem#1\endcsname}%
\let\auto@bib@innerbib\@empty
\bibitem [{\citenamefont {Kimel}\ \emph {et~al.}(2022)\citenamefont {Kimel}, \citenamefont {Zvezdin}, \citenamefont {Sharma}, \citenamefont {Shallcross}, \citenamefont {De~Sousa}, \citenamefont {Garc{\'\i}a-Mart{\'\i}n}, \citenamefont {Salvan}, \citenamefont {Hamrle}, \citenamefont {Stejskal}, \citenamefont {McCord} \emph {et~al.}}]{kimel20222022}%
  \BibitemOpen
  \bibfield  {author} {\bibinfo {author} {\bibfnamefont {A.}~\bibnamefont {Kimel}}, \bibinfo {author} {\bibfnamefont {A.}~\bibnamefont {Zvezdin}}, \bibinfo {author} {\bibfnamefont {S.}~\bibnamefont {Sharma}}, \bibinfo {author} {\bibfnamefont {S.}~\bibnamefont {Shallcross}}, \bibinfo {author} {\bibfnamefont {N.}~\bibnamefont {De~Sousa}}, \bibinfo {author} {\bibfnamefont {A.}~\bibnamefont {Garc{\'\i}a-Mart{\'\i}n}}, \bibinfo {author} {\bibfnamefont {G.}~\bibnamefont {Salvan}}, \bibinfo {author} {\bibfnamefont {J.}~\bibnamefont {Hamrle}}, \bibinfo {author} {\bibfnamefont {O.}~\bibnamefont {Stejskal}}, \bibinfo {author} {\bibfnamefont {J.}~\bibnamefont {McCord}}, \emph {et~al.},\ }\bibfield  {title} {\bibinfo {title} {The 2022 magneto-optics roadmap},\ }\href@noop {} {\bibfield  {journal} {\bibinfo  {journal} {Journal of Physics D: Applied Physics}\ }\textbf {\bibinfo {volume} {55}},\ \bibinfo {pages} {463003} (\bibinfo {year} {2022})}\BibitemShut {NoStop}%
\bibitem [{\citenamefont {{\'S}migaj}\ \emph {et~al.}(2010)\citenamefont {{\'S}migaj}, \citenamefont {Romero-Vivas}, \citenamefont {Gralak}, \citenamefont {Magdenko}, \citenamefont {Dagens},\ and\ \citenamefont {Vanwolleghem}}]{smigaj2010magneto}%
  \BibitemOpen
  \bibfield  {author} {\bibinfo {author} {\bibfnamefont {W.}~\bibnamefont {{\'S}migaj}}, \bibinfo {author} {\bibfnamefont {J.}~\bibnamefont {Romero-Vivas}}, \bibinfo {author} {\bibfnamefont {B.}~\bibnamefont {Gralak}}, \bibinfo {author} {\bibfnamefont {L.}~\bibnamefont {Magdenko}}, \bibinfo {author} {\bibfnamefont {B.}~\bibnamefont {Dagens}},\ and\ \bibinfo {author} {\bibfnamefont {M.}~\bibnamefont {Vanwolleghem}},\ }\bibfield  {title} {\bibinfo {title} {Magneto-optical circulator designed for operation in a uniform external magnetic field},\ }\href@noop {} {\bibfield  {journal} {\bibinfo  {journal} {Optics letters}\ }\textbf {\bibinfo {volume} {35}},\ \bibinfo {pages} {568} (\bibinfo {year} {2010})}\BibitemShut {NoStop}%
\bibitem [{\citenamefont {Inoue}\ \emph {et~al.}(2013)\citenamefont {Inoue}, \citenamefont {Levy},\ and\ \citenamefont {Baryshev}}]{inoue2013magnetophotonics}%
  \BibitemOpen
  \bibfield  {author} {\bibinfo {author} {\bibfnamefont {M.}~\bibnamefont {Inoue}}, \bibinfo {author} {\bibfnamefont {M.}~\bibnamefont {Levy}},\ and\ \bibinfo {author} {\bibfnamefont {A.~V.}\ \bibnamefont {Baryshev}},\ }\href@noop {} {\emph {\bibinfo {title} {Magnetophotonics: From theory to applications}}}\ (\bibinfo  {publisher} {Springer Science \& Business Media},\ \bibinfo {year} {2013})\BibitemShut {NoStop}%
\bibitem [{\citenamefont {Karki}\ \emph {et~al.}(2017)\citenamefont {Karki}, \citenamefont {Stenger}, \citenamefont {Pollick},\ and\ \citenamefont {Levy}}]{karki2017thin}%
  \BibitemOpen
  \bibfield  {author} {\bibinfo {author} {\bibfnamefont {D.}~\bibnamefont {Karki}}, \bibinfo {author} {\bibfnamefont {V.}~\bibnamefont {Stenger}}, \bibinfo {author} {\bibfnamefont {A.}~\bibnamefont {Pollick}},\ and\ \bibinfo {author} {\bibfnamefont {M.}~\bibnamefont {Levy}},\ }\bibfield  {title} {\bibinfo {title} {Thin-film magnetless faraday rotators for compact heterogeneous integrated optical isolators},\ }\href@noop {} {\bibfield  {journal} {\bibinfo  {journal} {Journal of Applied Physics}\ }\textbf {\bibinfo {volume} {121}} (\bibinfo {year} {2017})}\BibitemShut {NoStop}%
\bibitem [{\citenamefont {Ho}\ \emph {et~al.}(2018)\citenamefont {Ho}, \citenamefont {Im}, \citenamefont {Pae}, \citenamefont {Ri}, \citenamefont {Han},\ and\ \citenamefont {Herrmann}}]{ho2018switchable}%
  \BibitemOpen
  \bibfield  {author} {\bibinfo {author} {\bibfnamefont {K.-S.}\ \bibnamefont {Ho}}, \bibinfo {author} {\bibfnamefont {S.-J.}\ \bibnamefont {Im}}, \bibinfo {author} {\bibfnamefont {J.-S.}\ \bibnamefont {Pae}}, \bibinfo {author} {\bibfnamefont {C.-S.}\ \bibnamefont {Ri}}, \bibinfo {author} {\bibfnamefont {Y.-H.}\ \bibnamefont {Han}},\ and\ \bibinfo {author} {\bibfnamefont {J.}~\bibnamefont {Herrmann}},\ }\bibfield  {title} {\bibinfo {title} {Switchable plasmonic routers controlled by external magnetic fields by using magneto-plasmonic waveguides},\ }\href@noop {} {\bibfield  {journal} {\bibinfo  {journal} {Scientific reports}\ }\textbf {\bibinfo {volume} {8}},\ \bibinfo {pages} {1} (\bibinfo {year} {2018})}\BibitemShut {NoStop}%
\bibitem [{\citenamefont {Maccaferri}\ \emph {et~al.}(2016)\citenamefont {Maccaferri}, \citenamefont {Bergamini}, \citenamefont {Pancaldi}, \citenamefont {Schmidt}, \citenamefont {Kataja}, \citenamefont {Dijken}, \citenamefont {Zabala}, \citenamefont {Aizpurua},\ and\ \citenamefont {Vavassori}}]{maccaferri2016anisotropic}%
  \BibitemOpen
  \bibfield  {author} {\bibinfo {author} {\bibfnamefont {N.}~\bibnamefont {Maccaferri}}, \bibinfo {author} {\bibfnamefont {L.}~\bibnamefont {Bergamini}}, \bibinfo {author} {\bibfnamefont {M.}~\bibnamefont {Pancaldi}}, \bibinfo {author} {\bibfnamefont {M.~K.}\ \bibnamefont {Schmidt}}, \bibinfo {author} {\bibfnamefont {M.}~\bibnamefont {Kataja}}, \bibinfo {author} {\bibfnamefont {S.~v.}\ \bibnamefont {Dijken}}, \bibinfo {author} {\bibfnamefont {N.}~\bibnamefont {Zabala}}, \bibinfo {author} {\bibfnamefont {J.}~\bibnamefont {Aizpurua}},\ and\ \bibinfo {author} {\bibfnamefont {P.}~\bibnamefont {Vavassori}},\ }\bibfield  {title} {\bibinfo {title} {Anisotropic nanoantenna-based magnetoplasmonic crystals for highly enhanced and tunable magneto-optical activity},\ }\href@noop {} {\bibfield  {journal} {\bibinfo  {journal} {Nano letters}\ }\textbf {\bibinfo {volume} {16}},\ \bibinfo {pages} {2533} (\bibinfo {year} {2016})}\BibitemShut {NoStop}%
\bibitem [{\citenamefont {Fan}\ \emph {et~al.}(2019)\citenamefont {Fan}, \citenamefont {Nasir}, \citenamefont {Nicholls}, \citenamefont {Zayats},\ and\ \citenamefont {Podolskiy}}]{fan2019magneto}%
  \BibitemOpen
  \bibfield  {author} {\bibinfo {author} {\bibfnamefont {B.}~\bibnamefont {Fan}}, \bibinfo {author} {\bibfnamefont {M.~E.}\ \bibnamefont {Nasir}}, \bibinfo {author} {\bibfnamefont {L.~H.}\ \bibnamefont {Nicholls}}, \bibinfo {author} {\bibfnamefont {A.~V.}\ \bibnamefont {Zayats}},\ and\ \bibinfo {author} {\bibfnamefont {V.~A.}\ \bibnamefont {Podolskiy}},\ }\bibfield  {title} {\bibinfo {title} {Magneto-optical metamaterials: nonreciprocal transmission and faraday effect enhancement},\ }\href@noop {} {\bibfield  {journal} {\bibinfo  {journal} {Advanced Optical Materials}\ }\textbf {\bibinfo {volume} {7}},\ \bibinfo {pages} {1801420} (\bibinfo {year} {2019})}\BibitemShut {NoStop}%
\bibitem [{\citenamefont {Guo}\ \emph {et~al.}(2018)\citenamefont {Guo}, \citenamefont {Wu}, \citenamefont {Xue}, \citenamefont {Jiang}, \citenamefont {Sun}, \citenamefont {Li},\ and\ \citenamefont {Chen}}]{guo2018significant}%
  \BibitemOpen
  \bibfield  {author} {\bibinfo {author} {\bibfnamefont {Z.}~\bibnamefont {Guo}}, \bibinfo {author} {\bibfnamefont {F.}~\bibnamefont {Wu}}, \bibinfo {author} {\bibfnamefont {C.}~\bibnamefont {Xue}}, \bibinfo {author} {\bibfnamefont {H.}~\bibnamefont {Jiang}}, \bibinfo {author} {\bibfnamefont {Y.}~\bibnamefont {Sun}}, \bibinfo {author} {\bibfnamefont {Y.}~\bibnamefont {Li}},\ and\ \bibinfo {author} {\bibfnamefont {H.}~\bibnamefont {Chen}},\ }\bibfield  {title} {\bibinfo {title} {Significant enhancement of magneto-optical effect in one-dimensional photonic crystals with a magnetized epsilon-near-zero defect},\ }\href@noop {} {\bibfield  {journal} {\bibinfo  {journal} {Journal of Applied Physics}\ }\textbf {\bibinfo {volume} {124}} (\bibinfo {year} {2018})}\BibitemShut {NoStop}%
\bibitem [{\citenamefont {Xia}\ \emph {et~al.}(2022)\citenamefont {Xia}, \citenamefont {Ignatyeva}, \citenamefont {Liu}, \citenamefont {Qin}, \citenamefont {Kang}, \citenamefont {Yang}, \citenamefont {Chen}, \citenamefont {Duan}, \citenamefont {Deng}, \citenamefont {Long} \emph {et~al.}}]{xia2022enhancement}%
  \BibitemOpen
  \bibfield  {author} {\bibinfo {author} {\bibfnamefont {S.}~\bibnamefont {Xia}}, \bibinfo {author} {\bibfnamefont {D.~O.}\ \bibnamefont {Ignatyeva}}, \bibinfo {author} {\bibfnamefont {Q.}~\bibnamefont {Liu}}, \bibinfo {author} {\bibfnamefont {J.}~\bibnamefont {Qin}}, \bibinfo {author} {\bibfnamefont {T.}~\bibnamefont {Kang}}, \bibinfo {author} {\bibfnamefont {W.}~\bibnamefont {Yang}}, \bibinfo {author} {\bibfnamefont {Y.}~\bibnamefont {Chen}}, \bibinfo {author} {\bibfnamefont {H.}~\bibnamefont {Duan}}, \bibinfo {author} {\bibfnamefont {L.}~\bibnamefont {Deng}}, \bibinfo {author} {\bibfnamefont {D.}~\bibnamefont {Long}}, \emph {et~al.},\ }\bibfield  {title} {\bibinfo {title} {Enhancement of the faraday effect and magneto-optical figure of merit in all-dielectric metasurfaces},\ }\href@noop {} {\bibfield  {journal} {\bibinfo  {journal} {ACS Photonics}\ }\textbf {\bibinfo {volume} {9}},\ \bibinfo {pages} {1240} (\bibinfo {year} {2022})}\BibitemShut {NoStop}%
\bibitem [{\citenamefont {Barsukova}\ \emph {et~al.}(2019)\citenamefont {Barsukova}, \citenamefont {Musorin}, \citenamefont {Shorokhov},\ and\ \citenamefont {Fedyanin}}]{barsukova2019enhanced}%
  \BibitemOpen
  \bibfield  {author} {\bibinfo {author} {\bibfnamefont {M.}~\bibnamefont {Barsukova}}, \bibinfo {author} {\bibfnamefont {A.}~\bibnamefont {Musorin}}, \bibinfo {author} {\bibfnamefont {A.}~\bibnamefont {Shorokhov}},\ and\ \bibinfo {author} {\bibfnamefont {A.}~\bibnamefont {Fedyanin}},\ }\bibfield  {title} {\bibinfo {title} {Enhanced magneto-optical effects in hybrid ni-si metasurfaces},\ }\href@noop {} {\bibfield  {journal} {\bibinfo  {journal} {Apl Photonics}\ }\textbf {\bibinfo {volume} {4}} (\bibinfo {year} {2019})}\BibitemShut {NoStop}%
\bibitem [{\citenamefont {Christofi}\ \emph {et~al.}(2018)\citenamefont {Christofi}, \citenamefont {Kawaguchi}, \citenamefont {Al{\`u}},\ and\ \citenamefont {Khanikaev}}]{christofi2018giant}%
  \BibitemOpen
  \bibfield  {author} {\bibinfo {author} {\bibfnamefont {A.}~\bibnamefont {Christofi}}, \bibinfo {author} {\bibfnamefont {Y.}~\bibnamefont {Kawaguchi}}, \bibinfo {author} {\bibfnamefont {A.}~\bibnamefont {Al{\`u}}},\ and\ \bibinfo {author} {\bibfnamefont {A.~B.}\ \bibnamefont {Khanikaev}},\ }\bibfield  {title} {\bibinfo {title} {Giant enhancement of faraday rotation due to electromagnetically induced transparency in all-dielectric magneto-optical metasurfaces},\ }\href@noop {} {\bibfield  {journal} {\bibinfo  {journal} {Optics letters}\ }\textbf {\bibinfo {volume} {43}},\ \bibinfo {pages} {1838} (\bibinfo {year} {2018})}\BibitemShut {NoStop}%
\bibitem [{\citenamefont {Sadatgol}\ \emph {et~al.}(2016)\citenamefont {Sadatgol}, \citenamefont {Rahman}, \citenamefont {Forati}, \citenamefont {Levy},\ and\ \citenamefont {G{\"u}ney}}]{sadatgol2016enhanced}%
  \BibitemOpen
  \bibfield  {author} {\bibinfo {author} {\bibfnamefont {M.}~\bibnamefont {Sadatgol}}, \bibinfo {author} {\bibfnamefont {M.}~\bibnamefont {Rahman}}, \bibinfo {author} {\bibfnamefont {E.}~\bibnamefont {Forati}}, \bibinfo {author} {\bibfnamefont {M.}~\bibnamefont {Levy}},\ and\ \bibinfo {author} {\bibfnamefont {D.~{\"O}.}\ \bibnamefont {G{\"u}ney}},\ }\bibfield  {title} {\bibinfo {title} {Enhanced faraday rotation in hybrid magneto-optical metamaterial structure of bismuth-substituted-iron-garnet with embedded-gold-wires},\ }\href@noop {} {\bibfield  {journal} {\bibinfo  {journal} {Journal of Applied Physics}\ }\textbf {\bibinfo {volume} {119}} (\bibinfo {year} {2016})}\BibitemShut {NoStop}%
\bibitem [{\citenamefont {Royer}\ \emph {et~al.}(2020)\citenamefont {Royer}, \citenamefont {Varghese}, \citenamefont {Gamet}, \citenamefont {Neveu}, \citenamefont {Jourlin},\ and\ \citenamefont {Jamon}}]{royer2020enhancement}%
  \BibitemOpen
  \bibfield  {author} {\bibinfo {author} {\bibfnamefont {F.}~\bibnamefont {Royer}}, \bibinfo {author} {\bibfnamefont {B.}~\bibnamefont {Varghese}}, \bibinfo {author} {\bibfnamefont {E.}~\bibnamefont {Gamet}}, \bibinfo {author} {\bibfnamefont {S.}~\bibnamefont {Neveu}}, \bibinfo {author} {\bibfnamefont {Y.}~\bibnamefont {Jourlin}},\ and\ \bibinfo {author} {\bibfnamefont {D.}~\bibnamefont {Jamon}},\ }\bibfield  {title} {\bibinfo {title} {Enhancement of both faraday and kerr effects with an all-dielectric grating based on a magneto-optical nanocomposite material},\ }\href@noop {} {\bibfield  {journal} {\bibinfo  {journal} {ACS omega}\ }\textbf {\bibinfo {volume} {5}},\ \bibinfo {pages} {2886} (\bibinfo {year} {2020})}\BibitemShut {NoStop}%
\bibitem [{\citenamefont {Bsawmaii}\ \emph {et~al.}(2022)\citenamefont {Bsawmaii}, \citenamefont {Gamet}, \citenamefont {Neveu}, \citenamefont {Jamon},\ and\ \citenamefont {Royer}}]{bsawmaii2022magnetic}%
  \BibitemOpen
  \bibfield  {author} {\bibinfo {author} {\bibfnamefont {L.}~\bibnamefont {Bsawmaii}}, \bibinfo {author} {\bibfnamefont {E.}~\bibnamefont {Gamet}}, \bibinfo {author} {\bibfnamefont {S.}~\bibnamefont {Neveu}}, \bibinfo {author} {\bibfnamefont {D.}~\bibnamefont {Jamon}},\ and\ \bibinfo {author} {\bibfnamefont {F.}~\bibnamefont {Royer}},\ }\bibfield  {title} {\bibinfo {title} {Magnetic nanocomposite films with photo-patterned 1d grating on top enable giant magneto-optical intensity effects},\ }\href@noop {} {\bibfield  {journal} {\bibinfo  {journal} {Optical Materials Express}\ }\textbf {\bibinfo {volume} {12}},\ \bibinfo {pages} {513} (\bibinfo {year} {2022})}\BibitemShut {NoStop}%
\bibitem [{\citenamefont {Zograf}\ \emph {et~al.}(2021)\citenamefont {Zograf}, \citenamefont {Petrov}, \citenamefont {Makarov},\ and\ \citenamefont {Kivshar}}]{zograf2021all}%
  \BibitemOpen
  \bibfield  {author} {\bibinfo {author} {\bibfnamefont {G.~P.}\ \bibnamefont {Zograf}}, \bibinfo {author} {\bibfnamefont {M.~I.}\ \bibnamefont {Petrov}}, \bibinfo {author} {\bibfnamefont {S.~V.}\ \bibnamefont {Makarov}},\ and\ \bibinfo {author} {\bibfnamefont {Y.~S.}\ \bibnamefont {Kivshar}},\ }\bibfield  {title} {\bibinfo {title} {All-dielectric thermonanophotonics},\ }\href@noop {} {\bibfield  {journal} {\bibinfo  {journal} {Advances in Optics and Photonics}\ }\textbf {\bibinfo {volume} {13}},\ \bibinfo {pages} {643} (\bibinfo {year} {2021})}\BibitemShut {NoStop}%
\bibitem [{\citenamefont {Testa-Anta}\ \emph {et~al.}(2021)\citenamefont {Testa-Anta}, \citenamefont {Sousa-Castillo}, \citenamefont {L{\'o}pez-Ortega}, \citenamefont {Correa-Duarte}, \citenamefont {Garc{\'\i}a-Mart{\'\i}n}, \citenamefont {Vavassori},\ and\ \citenamefont {Salgueiri{\~n}o}}]{testa2021caging}%
  \BibitemOpen
  \bibfield  {author} {\bibinfo {author} {\bibfnamefont {M.}~\bibnamefont {Testa-Anta}}, \bibinfo {author} {\bibfnamefont {A.}~\bibnamefont {Sousa-Castillo}}, \bibinfo {author} {\bibfnamefont {A.}~\bibnamefont {L{\'o}pez-Ortega}}, \bibinfo {author} {\bibfnamefont {M.~A.}\ \bibnamefont {Correa-Duarte}}, \bibinfo {author} {\bibfnamefont {A.}~\bibnamefont {Garc{\'\i}a-Mart{\'\i}n}}, \bibinfo {author} {\bibfnamefont {P.}~\bibnamefont {Vavassori}},\ and\ \bibinfo {author} {\bibfnamefont {V.}~\bibnamefont {Salgueiri{\~n}o}},\ }\bibfield  {title} {\bibinfo {title} {A caging strategy for tuning the magneto-optical properties of cobalt ferrite using a single plasmonic nanoparticle},\ }\href@noop {} {\bibfield  {journal} {\bibinfo  {journal} {Journal of Materials Chemistry C}\ }\textbf {\bibinfo {volume} {9}},\ \bibinfo {pages} {5098} (\bibinfo {year} {2021})}\BibitemShut {NoStop}%
\bibitem [{\citenamefont {Kolmychek}\ \emph {et~al.}(2015)\citenamefont {Kolmychek}, \citenamefont {Shaimanov}, \citenamefont {Baryshev},\ and\ \citenamefont {Murzina}}]{kolmychek2015magneto}%
  \BibitemOpen
  \bibfield  {author} {\bibinfo {author} {\bibfnamefont {I.~A.}\ \bibnamefont {Kolmychek}}, \bibinfo {author} {\bibfnamefont {A.~N.}\ \bibnamefont {Shaimanov}}, \bibinfo {author} {\bibfnamefont {A.~V.}\ \bibnamefont {Baryshev}},\ and\ \bibinfo {author} {\bibfnamefont {T.~V.}\ \bibnamefont {Murzina}},\ }\bibfield  {title} {\bibinfo {title} {Magneto-optical response of two-dimensional magnetic plasmon structures based on gold nanodisks embedded in a ferrite garnet layer},\ }\href@noop {} {\bibfield  {journal} {\bibinfo  {journal} {JETP letters}\ }\textbf {\bibinfo {volume} {102}},\ \bibinfo {pages} {46} (\bibinfo {year} {2015})}\BibitemShut {NoStop}%
\bibitem [{\citenamefont {Ignatyeva}\ \emph {et~al.}(2020)\citenamefont {Ignatyeva}, \citenamefont {Karki}, \citenamefont {Voronov}, \citenamefont {Kozhaev}, \citenamefont {Krichevsky}, \citenamefont {Chernov}, \citenamefont {Levy},\ and\ \citenamefont {Belotelov}}]{ignatyeva2020all}%
  \BibitemOpen
  \bibfield  {author} {\bibinfo {author} {\bibfnamefont {D.~O.}\ \bibnamefont {Ignatyeva}}, \bibinfo {author} {\bibfnamefont {D.}~\bibnamefont {Karki}}, \bibinfo {author} {\bibfnamefont {A.~A.}\ \bibnamefont {Voronov}}, \bibinfo {author} {\bibfnamefont {M.~A.}\ \bibnamefont {Kozhaev}}, \bibinfo {author} {\bibfnamefont {D.~M.}\ \bibnamefont {Krichevsky}}, \bibinfo {author} {\bibfnamefont {A.~I.}\ \bibnamefont {Chernov}}, \bibinfo {author} {\bibfnamefont {M.}~\bibnamefont {Levy}},\ and\ \bibinfo {author} {\bibfnamefont {V.~I.}\ \bibnamefont {Belotelov}},\ }\bibfield  {title} {\bibinfo {title} {All-dielectric magnetic metasurface for advanced light control in dual polarizations combined with high-q resonances},\ }\href@noop {} {\bibfield  {journal} {\bibinfo  {journal} {Nature communications}\ }\textbf {\bibinfo {volume} {11}},\ \bibinfo {pages} {5487} (\bibinfo {year} {2020})}\BibitemShut {NoStop}%
\bibitem [{\citenamefont {Voronov}\ \emph {et~al.}(2020)\citenamefont {Voronov}, \citenamefont {Karki}, \citenamefont {Ignatyeva}, \citenamefont {Kozhaev}, \citenamefont {Levy},\ and\ \citenamefont {Belotelov}}]{voronov2020magneto}%
  \BibitemOpen
  \bibfield  {author} {\bibinfo {author} {\bibfnamefont {A.~A.}\ \bibnamefont {Voronov}}, \bibinfo {author} {\bibfnamefont {D.}~\bibnamefont {Karki}}, \bibinfo {author} {\bibfnamefont {D.~O.}\ \bibnamefont {Ignatyeva}}, \bibinfo {author} {\bibfnamefont {M.~A.}\ \bibnamefont {Kozhaev}}, \bibinfo {author} {\bibfnamefont {M.}~\bibnamefont {Levy}},\ and\ \bibinfo {author} {\bibfnamefont {V.~I.}\ \bibnamefont {Belotelov}},\ }\bibfield  {title} {\bibinfo {title} {Magneto-optics of subwavelength all-dielectric gratings},\ }\href@noop {} {\bibfield  {journal} {\bibinfo  {journal} {Optics Express}\ }\textbf {\bibinfo {volume} {28}},\ \bibinfo {pages} {17988} (\bibinfo {year} {2020})}\BibitemShut {NoStop}%
\bibitem [{\citenamefont {Bsawmaii}\ \emph {et~al.}(2020)\citenamefont {Bsawmaii}, \citenamefont {Gamet}, \citenamefont {Royer}, \citenamefont {Neveu},\ and\ \citenamefont {Jamon}}]{bsawmaii2020longitudinal}%
  \BibitemOpen
  \bibfield  {author} {\bibinfo {author} {\bibfnamefont {L.}~\bibnamefont {Bsawmaii}}, \bibinfo {author} {\bibfnamefont {E.}~\bibnamefont {Gamet}}, \bibinfo {author} {\bibfnamefont {F.}~\bibnamefont {Royer}}, \bibinfo {author} {\bibfnamefont {S.}~\bibnamefont {Neveu}},\ and\ \bibinfo {author} {\bibfnamefont {D.}~\bibnamefont {Jamon}},\ }\bibfield  {title} {\bibinfo {title} {Longitudinal magneto-optical effect enhancement with high transmission through a 1d all-dielectric resonant guided mode grating},\ }\href@noop {} {\bibfield  {journal} {\bibinfo  {journal} {Optics express}\ }\textbf {\bibinfo {volume} {28}},\ \bibinfo {pages} {8436} (\bibinfo {year} {2020})}\BibitemShut {NoStop}%
\bibitem [{\citenamefont {Pomozov}\ \emph {et~al.}(2022)\citenamefont {Pomozov}, \citenamefont {Maydykovskiy}, \citenamefont {Ketsko}, \citenamefont {Stognij},\ and\ \citenamefont {Murzina}}]{pomozov2022resonant}%
  \BibitemOpen
  \bibfield  {author} {\bibinfo {author} {\bibfnamefont {A.~R.}\ \bibnamefont {Pomozov}}, \bibinfo {author} {\bibfnamefont {A.~I.}\ \bibnamefont {Maydykovskiy}}, \bibinfo {author} {\bibfnamefont {V.~A.}\ \bibnamefont {Ketsko}}, \bibinfo {author} {\bibfnamefont {A.~I.}\ \bibnamefont {Stognij}},\ and\ \bibinfo {author} {\bibfnamefont {T.~V.}\ \bibnamefont {Murzina}},\ }\bibfield  {title} {\bibinfo {title} {Resonant magnetooptical effects in encapsulated 1d plasmonic crystals},\ }\href@noop {} {\bibfield  {journal} {\bibinfo  {journal} {Advanced Photonics Research}\ }\textbf {\bibinfo {volume} {3}},\ \bibinfo {pages} {2100329} (\bibinfo {year} {2022})}\BibitemShut {NoStop}%
\bibitem [{\citenamefont {Barsukova}\ \emph {et~al.}(2017)\citenamefont {Barsukova}, \citenamefont {Shorokhov}, \citenamefont {Musorin}, \citenamefont {Neshev}, \citenamefont {Kivshar},\ and\ \citenamefont {Fedyanin}}]{barsukova2017magneto}%
  \BibitemOpen
  \bibfield  {author} {\bibinfo {author} {\bibfnamefont {M.~G.}\ \bibnamefont {Barsukova}}, \bibinfo {author} {\bibfnamefont {A.~S.}\ \bibnamefont {Shorokhov}}, \bibinfo {author} {\bibfnamefont {A.~I.}\ \bibnamefont {Musorin}}, \bibinfo {author} {\bibfnamefont {D.~N.}\ \bibnamefont {Neshev}}, \bibinfo {author} {\bibfnamefont {Y.~S.}\ \bibnamefont {Kivshar}},\ and\ \bibinfo {author} {\bibfnamefont {A.~A.}\ \bibnamefont {Fedyanin}},\ }\bibfield  {title} {\bibinfo {title} {Magneto-optical response enhanced by mie resonances in nanoantennas},\ }\href@noop {} {\bibfield  {journal} {\bibinfo  {journal} {Acs Photonics}\ }\textbf {\bibinfo {volume} {4}},\ \bibinfo {pages} {2390} (\bibinfo {year} {2017})}\BibitemShut {NoStop}%
\bibitem [{\citenamefont {Chernyak}\ \emph {et~al.}(2024)\citenamefont {Chernyak}, \citenamefont {Musorin},\ and\ \citenamefont {Fedyanin}}]{chernyak2024faraday}%
  \BibitemOpen
  \bibfield  {author} {\bibinfo {author} {\bibfnamefont {A.}~\bibnamefont {Chernyak}}, \bibinfo {author} {\bibfnamefont {A.}~\bibnamefont {Musorin}},\ and\ \bibinfo {author} {\bibfnamefont {A.}~\bibnamefont {Fedyanin}},\ }\bibfield  {title} {\bibinfo {title} {Faraday rotation dynamics in hybrid magnetophotonic metasurfaces},\ }\href@noop {} {\bibfield  {journal} {\bibinfo  {journal} {Moscow University Physics Bulletin}\ }\textbf {\bibinfo {volume} {79}},\ \bibinfo {pages} {592} (\bibinfo {year} {2024})}\BibitemShut {NoStop}%
\bibitem [{\citenamefont {Zimnyakova}\ \emph {et~al.}(2021)\citenamefont {Zimnyakova}, \citenamefont {Ignatyeva}, \citenamefont {Karki}, \citenamefont {Voronov}, \citenamefont {Shaposhnikov}, \citenamefont {Berzhansky}, \citenamefont {Levy},\ and\ \citenamefont {Belotelov}}]{zimnyakova2021two}%
  \BibitemOpen
  \bibfield  {author} {\bibinfo {author} {\bibfnamefont {P.~E.}\ \bibnamefont {Zimnyakova}}, \bibinfo {author} {\bibfnamefont {D.~O.}\ \bibnamefont {Ignatyeva}}, \bibinfo {author} {\bibfnamefont {D.}~\bibnamefont {Karki}}, \bibinfo {author} {\bibfnamefont {A.~A.}\ \bibnamefont {Voronov}}, \bibinfo {author} {\bibfnamefont {A.~N.}\ \bibnamefont {Shaposhnikov}}, \bibinfo {author} {\bibfnamefont {V.~N.}\ \bibnamefont {Berzhansky}}, \bibinfo {author} {\bibfnamefont {M.}~\bibnamefont {Levy}},\ and\ \bibinfo {author} {\bibfnamefont {V.~I.}\ \bibnamefont {Belotelov}},\ }\bibfield  {title} {\bibinfo {title} {Two-dimensional array of iron-garnet nanocylinders supporting localized and lattice modes for the broadband boosted magneto-optics},\ }\href@noop {} {\bibfield  {journal} {\bibinfo  {journal} {Nanophotonics}\ }\textbf {\bibinfo {volume} {11}},\ \bibinfo {pages} {119} (\bibinfo {year} {2021})}\BibitemShut {NoStop}%
\bibitem [{\citenamefont {Krichevsky}\ \emph {et~al.}(2021)\citenamefont {Krichevsky}, \citenamefont {Xia}, \citenamefont {Mandrik}, \citenamefont {Ignatyeva}, \citenamefont {Bi},\ and\ \citenamefont {Belotelov}}]{krichevsky2021silicon}%
  \BibitemOpen
  \bibfield  {author} {\bibinfo {author} {\bibfnamefont {D.~M.}\ \bibnamefont {Krichevsky}}, \bibinfo {author} {\bibfnamefont {S.}~\bibnamefont {Xia}}, \bibinfo {author} {\bibfnamefont {M.~P.}\ \bibnamefont {Mandrik}}, \bibinfo {author} {\bibfnamefont {D.~O.}\ \bibnamefont {Ignatyeva}}, \bibinfo {author} {\bibfnamefont {L.}~\bibnamefont {Bi}},\ and\ \bibinfo {author} {\bibfnamefont {V.~I.}\ \bibnamefont {Belotelov}},\ }\bibfield  {title} {\bibinfo {title} {Silicon-based all-dielectric metasurface on an iron garnet film for efficient magneto-optical light modulation in near ir range},\ }\href@noop {} {\bibfield  {journal} {\bibinfo  {journal} {Nanomaterials}\ }\textbf {\bibinfo {volume} {11}},\ \bibinfo {pages} {2926} (\bibinfo {year} {2021})}\BibitemShut {NoStop}%
\bibitem [{\citenamefont {Chang}\ \emph {et~al.}(2012)\citenamefont {Chang}, \citenamefont {Jhu},\ and\ \citenamefont {Wu}}]{chang2012temperature}%
  \BibitemOpen
  \bibfield  {author} {\bibinfo {author} {\bibfnamefont {Y.-H.}\ \bibnamefont {Chang}}, \bibinfo {author} {\bibfnamefont {Y.-Y.}\ \bibnamefont {Jhu}},\ and\ \bibinfo {author} {\bibfnamefont {C.-J.}\ \bibnamefont {Wu}},\ }\bibfield  {title} {\bibinfo {title} {Temperature dependence of defect mode in a defective photonic crystal},\ }\href@noop {} {\bibfield  {journal} {\bibinfo  {journal} {Optics Communications}\ }\textbf {\bibinfo {volume} {285}},\ \bibinfo {pages} {1501} (\bibinfo {year} {2012})}\BibitemShut {NoStop}%
\bibitem [{\citenamefont {Zhang}\ \emph {et~al.}(2019)\citenamefont {Zhang}, \citenamefont {Wei}, \citenamefont {Rukhlenko}, \citenamefont {Chen},\ and\ \citenamefont {Zhu}}]{zhang2019electrically}%
  \BibitemOpen
  \bibfield  {author} {\bibinfo {author} {\bibfnamefont {J.}~\bibnamefont {Zhang}}, \bibinfo {author} {\bibfnamefont {X.}~\bibnamefont {Wei}}, \bibinfo {author} {\bibfnamefont {I.~D.}\ \bibnamefont {Rukhlenko}}, \bibinfo {author} {\bibfnamefont {H.-T.}\ \bibnamefont {Chen}},\ and\ \bibinfo {author} {\bibfnamefont {W.}~\bibnamefont {Zhu}},\ }\bibfield  {title} {\bibinfo {title} {Electrically tunable metasurface with independent frequency and amplitude modulations},\ }\href@noop {} {\bibfield  {journal} {\bibinfo  {journal} {Acs Photonics}\ }\textbf {\bibinfo {volume} {7}},\ \bibinfo {pages} {265} (\bibinfo {year} {2019})}\BibitemShut {NoStop}%
\bibitem [{\citenamefont {Li}\ \emph {et~al.}(2020)\citenamefont {Li}, \citenamefont {Lin}, \citenamefont {Guo}, \citenamefont {Sun}, \citenamefont {Xiao},\ and\ \citenamefont {Zhou}}]{li2020tunable}%
  \BibitemOpen
  \bibfield  {author} {\bibinfo {author} {\bibfnamefont {Y.}~\bibnamefont {Li}}, \bibinfo {author} {\bibfnamefont {J.}~\bibnamefont {Lin}}, \bibinfo {author} {\bibfnamefont {H.}~\bibnamefont {Guo}}, \bibinfo {author} {\bibfnamefont {W.}~\bibnamefont {Sun}}, \bibinfo {author} {\bibfnamefont {S.}~\bibnamefont {Xiao}},\ and\ \bibinfo {author} {\bibfnamefont {L.}~\bibnamefont {Zhou}},\ }\bibfield  {title} {\bibinfo {title} {A tunable metasurface with switchable functionalities: from perfect transparency to perfect absorption},\ }\href@noop {} {\bibfield  {journal} {\bibinfo  {journal} {Advanced optical materials}\ }\textbf {\bibinfo {volume} {8}},\ \bibinfo {pages} {1901548} (\bibinfo {year} {2020})}\BibitemShut {NoStop}%
\bibitem [{\citenamefont {Henriksson}\ \emph {et~al.}(2024)\citenamefont {Henriksson}, \citenamefont {Gabbani}, \citenamefont {Petrucci}, \citenamefont {Garoli}, \citenamefont {Pineider},\ and\ \citenamefont {Maccaferri}}]{henriksson2024probing}%
  \BibitemOpen
  \bibfield  {author} {\bibinfo {author} {\bibfnamefont {N.}~\bibnamefont {Henriksson}}, \bibinfo {author} {\bibfnamefont {A.}~\bibnamefont {Gabbani}}, \bibinfo {author} {\bibfnamefont {G.}~\bibnamefont {Petrucci}}, \bibinfo {author} {\bibfnamefont {D.}~\bibnamefont {Garoli}}, \bibinfo {author} {\bibfnamefont {F.}~\bibnamefont {Pineider}},\ and\ \bibinfo {author} {\bibfnamefont {N.}~\bibnamefont {Maccaferri}},\ }\bibfield  {title} {\bibinfo {title} {Probing temperature changes using nonradiative processes in hyperbolic meta-antennas},\ }\href@noop {} {\bibfield  {journal} {\bibinfo  {journal} {ACS Applied Optical Materials}\ }\textbf {\bibinfo {volume} {2}},\ \bibinfo {pages} {2469} (\bibinfo {year} {2024})}\BibitemShut {NoStop}%
\bibitem [{\citenamefont {St{\"o}hr}\ and\ \citenamefont {Siegmann}(2006)}]{stohr2006magnetism}%
  \BibitemOpen
  \bibfield  {author} {\bibinfo {author} {\bibfnamefont {J.}~\bibnamefont {St{\"o}hr}}\ and\ \bibinfo {author} {\bibfnamefont {H.~C.}\ \bibnamefont {Siegmann}},\ }\bibfield  {title} {\bibinfo {title} {Magnetism},\ }\href@noop {} {\bibfield  {journal} {\bibinfo  {journal} {Solid-State Sciences. Springer, Berlin, Heidelberg}\ }\textbf {\bibinfo {volume} {5}},\ \bibinfo {pages} {236} (\bibinfo {year} {2006})}\BibitemShut {NoStop}%
\bibitem [{\citenamefont {Tang}\ \emph {et~al.}(2024)\citenamefont {Tang}, \citenamefont {Zhang}, \citenamefont {Xiao}, \citenamefont {Qin}, \citenamefont {He}, \citenamefont {Liu}, \citenamefont {Liao},\ and\ \citenamefont {Yu}}]{tang2024giant}%
  \BibitemOpen
  \bibfield  {author} {\bibinfo {author} {\bibfnamefont {Q.}~\bibnamefont {Tang}}, \bibinfo {author} {\bibfnamefont {D.}~\bibnamefont {Zhang}}, \bibinfo {author} {\bibfnamefont {S.}~\bibnamefont {Xiao}}, \bibinfo {author} {\bibfnamefont {M.}~\bibnamefont {Qin}}, \bibinfo {author} {\bibfnamefont {J.}~\bibnamefont {He}}, \bibinfo {author} {\bibfnamefont {T.}~\bibnamefont {Liu}}, \bibinfo {author} {\bibfnamefont {Q.}~\bibnamefont {Liao}},\ and\ \bibinfo {author} {\bibfnamefont {T.}~\bibnamefont {Yu}},\ }\bibfield  {title} {\bibinfo {title} {Giant enhancement of the transverse magneto-optical kerr effect in etchless bismuth-substituted yttrium iron garnet empowered by quasi-bound states in the continuum},\ }\href@noop {} {\bibfield  {journal} {\bibinfo  {journal} {arXiv preprint arXiv:2409.05121}\ } (\bibinfo {year} {2024})}\BibitemShut {NoStop}%
\bibitem [{\citenamefont {Darki}\ \emph {et~al.}(2016)\citenamefont {Darki}, \citenamefont {Nezhad},\ and\ \citenamefont {Firouzeh}}]{darki2016temporal}%
  \BibitemOpen
  \bibfield  {author} {\bibinfo {author} {\bibfnamefont {B.~S.}\ \bibnamefont {Darki}}, \bibinfo {author} {\bibfnamefont {A.~Z.}\ \bibnamefont {Nezhad}},\ and\ \bibinfo {author} {\bibfnamefont {Z.~H.}\ \bibnamefont {Firouzeh}},\ }\bibfield  {title} {\bibinfo {title} {Temporal coupled mode analysis of one-dimensional magneto-photonic crystals with cavity structures},\ }\href@noop {} {\bibfield  {journal} {\bibinfo  {journal} {Journal of Magnetism and Magnetic Materials}\ }\textbf {\bibinfo {volume} {419}},\ \bibinfo {pages} {84} (\bibinfo {year} {2016})}\BibitemShut {NoStop}%
\bibitem [{\citenamefont {Zvezdin}\ and\ \citenamefont {Kotov}(1997)}]{zvezdin1997modern}%
  \BibitemOpen
  \bibfield  {author} {\bibinfo {author} {\bibfnamefont {A.~K.}\ \bibnamefont {Zvezdin}}\ and\ \bibinfo {author} {\bibfnamefont {V.~A.}\ \bibnamefont {Kotov}},\ }\href@noop {} {\emph {\bibinfo {title} {Modern magnetooptics and magnetooptical materials}}}\ (\bibinfo  {publisher} {CRC Press},\ \bibinfo {year} {1997})\BibitemShut {NoStop}%
\bibitem [{\citenamefont {Rizal}\ \emph {et~al.}(2021)\citenamefont {Rizal}, \citenamefont {Manera}, \citenamefont {Ignatyeva}, \citenamefont {Mej{\'\i}a-Salazar}, \citenamefont {Rella}, \citenamefont {Belotelov}, \citenamefont {Pineider},\ and\ \citenamefont {Maccaferri}}]{rizal2021magnetophotonics}%
  \BibitemOpen
  \bibfield  {author} {\bibinfo {author} {\bibfnamefont {C.}~\bibnamefont {Rizal}}, \bibinfo {author} {\bibfnamefont {M.~G.}\ \bibnamefont {Manera}}, \bibinfo {author} {\bibfnamefont {D.~O.}\ \bibnamefont {Ignatyeva}}, \bibinfo {author} {\bibfnamefont {J.~R.}\ \bibnamefont {Mej{\'\i}a-Salazar}}, \bibinfo {author} {\bibfnamefont {R.}~\bibnamefont {Rella}}, \bibinfo {author} {\bibfnamefont {V.~I.}\ \bibnamefont {Belotelov}}, \bibinfo {author} {\bibfnamefont {F.}~\bibnamefont {Pineider}},\ and\ \bibinfo {author} {\bibfnamefont {N.}~\bibnamefont {Maccaferri}},\ }\bibfield  {title} {\bibinfo {title} {Magnetophotonics for sensing and magnetometry toward industrial applications},\ }\href@noop {} {\bibfield  {journal} {\bibinfo  {journal} {Journal of Applied Physics}\ }\textbf {\bibinfo {volume} {130}} (\bibinfo {year} {2021})}\BibitemShut {NoStop}%
\bibitem [{\citenamefont {Tang}\ \emph {et~al.}(2023)\citenamefont {Tang}, \citenamefont {Zhang}, \citenamefont {Liu}, \citenamefont {Liu}, \citenamefont {Liao}, \citenamefont {He}, \citenamefont {Xiao},\ and\ \citenamefont {Yu}}]{tang2023enhancing}%
  \BibitemOpen
  \bibfield  {author} {\bibinfo {author} {\bibfnamefont {Q.}~\bibnamefont {Tang}}, \bibinfo {author} {\bibfnamefont {D.}~\bibnamefont {Zhang}}, \bibinfo {author} {\bibfnamefont {T.}~\bibnamefont {Liu}}, \bibinfo {author} {\bibfnamefont {W.}~\bibnamefont {Liu}}, \bibinfo {author} {\bibfnamefont {Q.}~\bibnamefont {Liao}}, \bibinfo {author} {\bibfnamefont {J.}~\bibnamefont {He}}, \bibinfo {author} {\bibfnamefont {S.}~\bibnamefont {Xiao}},\ and\ \bibinfo {author} {\bibfnamefont {T.}~\bibnamefont {Yu}},\ }\bibfield  {title} {\bibinfo {title} {Enhancing faraday and kerr rotations based on the toroidal dipole mode in an all-dielectric magneto-optical metasurface},\ }\href@noop {} {\bibfield  {journal} {\bibinfo  {journal} {Optics Letters}\ }\textbf {\bibinfo {volume} {48}},\ \bibinfo {pages} {3451} (\bibinfo {year} {2023})}\BibitemShut {NoStop}%
\bibitem [{\citenamefont {Borovkova}\ \emph {et~al.}(2019)\citenamefont {Borovkova}, \citenamefont {Ignatyeva}, \citenamefont {Sekatskii}, \citenamefont {Karabchevsky},\ and\ \citenamefont {Belotelov}}]{borovkova2019high}%
  \BibitemOpen
  \bibfield  {author} {\bibinfo {author} {\bibfnamefont {O.}~\bibnamefont {Borovkova}}, \bibinfo {author} {\bibfnamefont {D.}~\bibnamefont {Ignatyeva}}, \bibinfo {author} {\bibfnamefont {S.}~\bibnamefont {Sekatskii}}, \bibinfo {author} {\bibfnamefont {A.}~\bibnamefont {Karabchevsky}},\ and\ \bibinfo {author} {\bibfnamefont {V.}~\bibnamefont {Belotelov}},\ }\bibfield  {title} {\bibinfo {title} {High-q surface electromagnetic wave resonance excitation in magnetophotonic crystals for supersensitive detection of weak light absorption in the near-infrared},\ }\href@noop {} {\bibfield  {journal} {\bibinfo  {journal} {Photonics Research}\ }\textbf {\bibinfo {volume} {8}},\ \bibinfo {pages} {57} (\bibinfo {year} {2019})}\BibitemShut {NoStop}%
\bibitem [{\citenamefont {Merzlikin}\ \emph {et~al.}(2018)\citenamefont {Merzlikin}, \citenamefont {Kuznetsov},\ and\ \citenamefont {Baryshev}}]{merzlikin2018magneto}%
  \BibitemOpen
  \bibfield  {author} {\bibinfo {author} {\bibfnamefont {A.~M.}\ \bibnamefont {Merzlikin}}, \bibinfo {author} {\bibfnamefont {E.~V.}\ \bibnamefont {Kuznetsov}},\ and\ \bibinfo {author} {\bibfnamefont {A.~V.}\ \bibnamefont {Baryshev}},\ }\bibfield  {title} {\bibinfo {title} {Magneto-optical device based on polarization sensitivity for perspective biosensing applications},\ }\href@noop {} {\bibfield  {journal} {\bibinfo  {journal} {IEEE Sensors Journal}\ }\textbf {\bibinfo {volume} {18}},\ \bibinfo {pages} {5732} (\bibinfo {year} {2018})}\BibitemShut {NoStop}%
\bibitem [{\citenamefont {Danilin}\ \emph {et~al.}(2021)\citenamefont {Danilin}, \citenamefont {Slinkov}, \citenamefont {Lobanov}, \citenamefont {Min’kov},\ and\ \citenamefont {Bilenko}}]{danilin2021magneto}%
  \BibitemOpen
  \bibfield  {author} {\bibinfo {author} {\bibfnamefont {A.}~\bibnamefont {Danilin}}, \bibinfo {author} {\bibfnamefont {G.}~\bibnamefont {Slinkov}}, \bibinfo {author} {\bibfnamefont {V.}~\bibnamefont {Lobanov}}, \bibinfo {author} {\bibfnamefont {K.}~\bibnamefont {Min’kov}},\ and\ \bibinfo {author} {\bibfnamefont {I.}~\bibnamefont {Bilenko}},\ }\bibfield  {title} {\bibinfo {title} {Magneto-optical effects in a high-q whispering-gallery-mode resonator with a large verdet constant},\ }\href@noop {} {\bibfield  {journal} {\bibinfo  {journal} {Optics Letters}\ }\textbf {\bibinfo {volume} {46}},\ \bibinfo {pages} {2509} (\bibinfo {year} {2021})}\BibitemShut {NoStop}%
\bibitem [{\citenamefont {Ignatyeva}\ \emph {et~al.}(2022)\citenamefont {Ignatyeva}, \citenamefont {Krichevsky}, \citenamefont {Belotelov}, \citenamefont {Royer}, \citenamefont {Dash},\ and\ \citenamefont {Levy}}]{ignatyeva2022all}%
  \BibitemOpen
  \bibfield  {author} {\bibinfo {author} {\bibfnamefont {D.~O.}\ \bibnamefont {Ignatyeva}}, \bibinfo {author} {\bibfnamefont {D.~M.}\ \bibnamefont {Krichevsky}}, \bibinfo {author} {\bibfnamefont {V.~I.}\ \bibnamefont {Belotelov}}, \bibinfo {author} {\bibfnamefont {F.}~\bibnamefont {Royer}}, \bibinfo {author} {\bibfnamefont {S.}~\bibnamefont {Dash}},\ and\ \bibinfo {author} {\bibfnamefont {M.}~\bibnamefont {Levy}},\ }\bibfield  {title} {\bibinfo {title} {All-dielectric magneto-photonic metasurfaces},\ }\href@noop {} {\bibfield  {journal} {\bibinfo  {journal} {Journal of Applied Physics}\ }\textbf {\bibinfo {volume} {132}} (\bibinfo {year} {2022})}\BibitemShut {NoStop}%
\bibitem [{\citenamefont {Falek}\ \emph {et~al.}(2021)\citenamefont {Falek}, \citenamefont {Katiyi}, \citenamefont {Greenberg},\ and\ \citenamefont {Karabchevsky}}]{falek2021chip}%
  \BibitemOpen
  \bibfield  {author} {\bibinfo {author} {\bibfnamefont {E.}~\bibnamefont {Falek}}, \bibinfo {author} {\bibfnamefont {A.}~\bibnamefont {Katiyi}}, \bibinfo {author} {\bibfnamefont {Y.}~\bibnamefont {Greenberg}},\ and\ \bibinfo {author} {\bibfnamefont {A.}~\bibnamefont {Karabchevsky}},\ }\bibfield  {title} {\bibinfo {title} {On-chip metasurface-on-facets for ultra-high transmission through waveguides in near-infrared},\ }\href@noop {} {\bibfield  {journal} {\bibinfo  {journal} {Advanced Optical Materials}\ }\textbf {\bibinfo {volume} {9}},\ \bibinfo {pages} {2100130} (\bibinfo {year} {2021})}\BibitemShut {NoStop}%
\bibitem [{\citenamefont {Franta}\ \emph {et~al.}(2017)\citenamefont {Franta}, \citenamefont {Dubroka}, \citenamefont {Wang}, \citenamefont {Giglia}, \citenamefont {Voh{\'a}nka}, \citenamefont {Franta},\ and\ \citenamefont {Ohl{\'\i}dal}}]{franta2017temperature}%
  \BibitemOpen
  \bibfield  {author} {\bibinfo {author} {\bibfnamefont {D.}~\bibnamefont {Franta}}, \bibinfo {author} {\bibfnamefont {A.}~\bibnamefont {Dubroka}}, \bibinfo {author} {\bibfnamefont {C.}~\bibnamefont {Wang}}, \bibinfo {author} {\bibfnamefont {A.}~\bibnamefont {Giglia}}, \bibinfo {author} {\bibfnamefont {J.}~\bibnamefont {Voh{\'a}nka}}, \bibinfo {author} {\bibfnamefont {P.}~\bibnamefont {Franta}},\ and\ \bibinfo {author} {\bibfnamefont {I.}~\bibnamefont {Ohl{\'\i}dal}},\ }\bibfield  {title} {\bibinfo {title} {Temperature-dependent dispersion model of float zone crystalline silicon},\ }\href@noop {} {\bibfield  {journal} {\bibinfo  {journal} {Applied Surface Science}\ }\textbf {\bibinfo {volume} {421}},\ \bibinfo {pages} {405} (\bibinfo {year} {2017})}\BibitemShut {NoStop}%
\end{thebibliography}%

\end{document}